\documentclass[11pt,twoside]{article}
\usepackage{aaspp4}
\renewcommand{\maketitle}{}

\usepackage[T1]{fontenc}
\usepackage{amsmath}
\usepackage{graphics}

\makeatletter

\newcommand{\LyX}{L\kern-.1667em\lower.25em\hbox{Y}\kern-.125emX\@}
\newcommand{\noun}[1]{\textsc{#1}}
\let\SF@@footnote\footnote
\def\footnote{\ifx\protect\@typeset@protect
    \expandafter\SF@@footnote
  \else
    \expandafter\SF@gobble@opt
  \fi
}
\expandafter\def\csname SF@gobble@opt \endcsname{\@ifnextchar[%]
  \SF@gobble@twobracket
  \@gobble
}
\edef\SF@gobble@opt{\noexpand\protect
  \expandafter\noexpand\csname SF@gobble@opt \endcsname}
\def\SF@gobble@twobracket[#1]#2{}

\makeatother

\begin{document}

\newcommand{\as}{''}
 
\newcommand{\Mbh}{M_{\bullet }}

\newcommand{\Mo}{M_{\odot }}
 
\newcommand{\Ro}{R_{\odot }}

\newcommand{\SgrA}{\mathrm{Sgr}\, \mathrm{A}^{\star }}

\newcommand{\kmag}{^{\mathrm{m}}}
 
\newcommand{\HeI}{He{\sc i}/H{\sc i}\, }

\hfill{}ApJ accepted\\

\title{The distribution of stars near the super-massive black hole in the Galactic
Center}

\author{Tal Alexander}

\maketitle
\affil{Institute for Advanced Study, Olden Lane, Princeton, NJ 08540} 

\authoremail{tal@ias.edu}

\lefthead{T. Alexander} 

\righthead{The distribution of stars near the Black Hole in the Galactic Center}

\begin{abstract}
We analyze three sets of infrared star counts in the inner \( \sim 0.5 \)~pc
of the Galactic Center. We perform statistical tests on the star counts and
model in detail the extinction field and the effects of dwarf--giant collisions
on the luminosity function. We find that both the star counts and the depletion
of the brightest stars in the inner \( \sim 0.05 \) pc can be explained by
a \( n\propto r^{-\alpha } \) stellar cusp with \( \alpha  \) in the range
\( 3/2 \) to \( 7/4 \), in which the envelopes of the brightest giants are
destroyed by stellar collisions. Such a cusp is consistent with the Bahcall-Wolf
solution for the distribution of stars that have undergone two-body relaxation
around a black hole. We show that systematic uncertainties due to variable extinction
and unrelaxed stars are probably small, but deeper star counts are required
to confirm these results. We estimate that the tidal disruption rate of cusp
stars by the black hole is \( \mathrm{few}\times 10^{-5}\, \mathrm{yr}^{-1} \).

\end{abstract}

\section{Introduction}

The Galactic Center (GC) offers a unique opportunity to study the dynamical
effects of a super-massive black hole (BH) on the stellar population in its
immediate vicinity. Dynamical models of the evolution of such a system generically
predict the formation of a \emph{stellar cusp} with a characteristic slope that
reflects its formation history (Bahcall \& Wolf \cite{BW76}, \cite{BW77};
Young \cite{Young80}; Lee \& Goodman \cite{LG89}; Quinlan, Hernquist \& Sigurdsson
\cite{Quinlan95}). Although the existence of super-massive BHs in the center
of normal galaxies now appears to be the rule rather than the exception (Magorrian
et al. \cite{Magorrian98}), the observed central light distributions do not
seem to be correlated with the mass of the central black hole (Kormendy \& Richstone
\cite{KR95}; but see Magorrian et al. \cite{Magorrian98}). It appears that
even on the smallest resolved scales of about \( 0.1\as  \), the light distribution
still reflects the overall galactic dynamics, and that the effects of the BH
on the spatial distribution of the stars may be confined to yet smaller scales.
Even with better resolution, it will be very hard to observe individual stars
in the dense galactic cores, and the properties of the stellar population will
have to be inferred from the projected integrated light, which is completely
dominated by giants and supergiants. These rare stars may not be representative
of the general stellar population. Furthermore, their light-emitting envelopes
could be affected by various processes in the extreme environment inside a BH
cusp. The situation is further complicated by the fact that galactic centers
often display cusps in the light distribution on scales that are much too large
to be related to the presence of a super-massive BH in the core. Obviously,
other dynamical processes can also give rise to stellar cusps, so it is not
possible to claim that a central stellar cusp \emph{must} be related to the
presence of a BH (Phinney \cite{Phinney89}). 

This is to be contrasted with the situation in the GC, where precise measurements
of the apparent motions and radial velocities of individual stars have already
made it possible to weigh the central BH dynamically, independently of the light
distribution (Genzel et al. \cite{Genzel97}; Ghez et al. \cite{Ghez98}). This
makes it legitimate to interpret the stellar distribution very near the center
in the context of BH formation scenarios. These observations also yielded deep
infrared star counts, which extend the stellar census to stars far less luminous
and much more numerous than supergiants. However, the step from the observed
number counts to the actual stellar distribution is far from trivial. Observations
show that the stellar population of GC is a mixture of old and young stars,
which are probably the products of episodic star-formation throughout the GC
history (Genzel, Hollenbach \& Townes \cite{GHT94}; Serabyn \& Morris \cite{SM96}).
Some of the young and very luminous stars are not dynamically relaxed (Krabbe
et al. \cite{Krabbe95}; Genzel et al. \cite{Genzel96}). Spatial variations
in the spectral properties of the stellar population (Sellgren et al. \cite{Sellgren90};
Krabbe et al. \cite{Krabbe95}; Genzel et al. \cite{Genzel96}), and possible
spatial gradients in the extinction (Blum et al. \cite{Blum96}; Davidge \cite{Davidge98}),
could further complicate the interpretation of the star counts. The necessary
first step towards testing dynamical models in the inner GC is to establish
what is the stellar distribution there.

The evidence for a stellar cusp within the cusp radius of the GC (roughly, the
radius enclosing a stellar mass comparable to the BH mass, which is \( \sim 3 \)
pc in the GC) has so far been inconclusive or negative. The mass models that
are currently adopted in stellar proper motion studies of the GC do not have
an inner stellar cusp, but assume a smoothed isothermal distribution with a
core radius of \( 0.4 \) pc (\( 1\as =0.04 \) pc in the GC for a solar galactocentric
distance of 8 kpc (Carney et al. \cite{Carney95}). In this work we re-examine
the question of whether there is a stellar cusp in the inner \( 10\as  \) of
the GC. 

This study is motivated by several recent developments. Progress in extending
infrared observations of the GC to fainter magnitudes and higher angular resolutions
(Genzel et al. \cite{Genzel97}; Ghez et al. \cite{Ghez98}) and to larger samples
with multi-color information (Blum et al. \cite{Blum96}; Davidge et al. \cite{Davidge97};
Davidge \cite{Davidge98}) yielded a wealth of information on the distribution
of stars in the GC. In addition, these deep infrared counts in the GC, together
with infrared and optical star counts in Baade's window (Tiede et al. \cite{Tiede95};
Holtzman et al. \cite{Holtzman98}), now make it possible to reliably model
the infrared stellar luminosity function in the GC over a large luminosity range
(Alexander \& Sternberg \cite{AS99}). However, the high angular resolution
studies are primarily focused on tracking stellar proper motions for weighing
the BH, while the multi-color studies are focused on studying the spectral properties
of the stellar population. Consequently, the star counts, which are a by-product
of these studies, have not yet been analyzed in detail, and the additional information
supplied by the luminosity function has not been taken into account. Finally,
recent high resolution deep imaging of the central few arcseconds revealed an
apparent over-density of faint stars in the inner \( 1\as  \) relative to the
immediate surroundings, the so-called ``\( \SgrA  \) cluster'' (Genzel et
al. \cite{Genzel97}; Ghez et al. \cite{Ghez98}). This raises the question
whether a distinct stellar cluster with an atypically faint stellar population
could have possibly formed in the inner \( 1\as  \) or else migrated there
from farther away. 

It this study we analyze the observed stellar distribution in three published
data sets (\S\ref{sec: obs}): Keck data from Ghez et al. (\cite{Ghez98}),
OSIRIS data from Blum et al. (\cite{Blum96}) and SHARP+3D data from Genzel
et al. (\cite{Genzel96}) and Eckart \& Genzel (\cite{EG97}). As we will argue
below, these data point to a simpler explanation for the \( \SgrA  \) cluster,
namely that it is the tip of an underlying stellar cusp that smoothly rises
throughout the inner \( 10\as  \), and that the absence of luminous stars there
is due to a combination of envelope-destroying stellar collisions and projection
effects rather than an intrinsic change in the properties of the stellar population.
We will further show that the shape of the cusp agrees with that predicted by
the Bahcall-Wolf solution for the distribution of intermediate mass stars in
a multi-mass system that has undergone two-body relaxation around a BH (Bahcall
\& Wolf \cite{BW76}, \cite{BW77}; Murphy, Cohn \& Durisen \cite{MCD91}),
as appears to be the case in the inner GC.

This paper is organized as follows. In \S\ref{sec:stars} we describe the relevant
properties of stars in the GC and present the data that are used in our analysis.
In \S\ref{sec:stat} we describe the stellar density distributions that we use
to model the data and present the results of the statistical analysis of the
counts. Some anomalies in the counts are further discussed in Appendix~\ref{sec: ring}.
In \S\ref{sec:Ak} we investigate the effects of extinction on the star counts.
Readers who are mainly interested in the dynamical results of this work may
prefer to skip this section and proceed to \S\ref{sec:coll}, where we discuss
the effects of stellar collisions on the depletion of the brightest stars. The
orbitally-averaged collision rate is derived in Appendix~\ref{sec:orbit}. We
discuss the results in \S\ref{sec:discussion} and summarize our conclusions
in \S\ref{sec:summary}.

\section{Stars in the Galactic Center}

\label{sec:stars}

\subsection{The stellar population and stellar distribution}

\label{sec:pop}

The stellar population in the GC is a mixture of old Galactic Bulge stars and
younger stars that were produced in various star-formation episodes during the
lifetime of the Galaxy (Genzel, Hollenbach \& Townes \cite{GHT94}; Serabyn
\& Morris \cite{SM96}). About 80\% of the stars observed in the central parsec
down to \( K=14.5\kmag  \) are identified as late-type giants and supergiants
by their pronounced CO absorption. Late-type stars at the magnitude range \( 11\kmag \la K\la 14.5\kmag  \)
are K and M giants, which correspond to a mass range of \( 1 \)--\( 7\Mo  \)
and ages \( >\mathrm{few}\times 10^{9} \) yr. Late-type stars at the magnitude
range \( 9\kmag \la K\la 12\kmag  \) are probably asymptotic giant branch (AGB)
stars in the mass range \( 2 \)--\( 8\Mo  \) and ages of \( 1 \)--\( \mathrm{few}\times 10^{9} \)
yr. The most luminous late-type star, IRS7 (\( K=6.5\kmag  \)), is a massive
\( 15 \)--\( 25\Mo  \) supergiant with an age of no more than \( 10^{7} \)
yr. The late-type giants are dynamically relaxed and follow the galactic rotation
(Genzel at al \cite{Genzel96}).

The remaining \( \sim 20\% \) of the \( K<14.5\kmag  \) stars in the central
parsec appear to be recently formed stars, either exhibiting strong \noun{}\HeI\noun{ }emission
or a steep featureless reddened continuum (Krabbe et al. \cite{Krabbe95}; Genzel
et al. \cite{Genzel96}). Two extended red objects, which may be recently formed
stars still embedded in a dust shell, are also observed (Ott, Eckart \& Genzel
\cite{Ott98}). The \HeI stars are massive, hot and very luminous stars with
huge mass loss rates, which are probably undergoing a short lived phase lasting
less than \( 10^{5} \) yr, on their way to becoming classical Wolf-Rayet stars.
In this case, their progenitors must be \( \ga 20\Mo  \) stars with main-sequence
lifetimes shorter than \( 10^{7} \) yr. The \( \sim 20 \) \HeI stars in the
inner few arcseconds supply more than 50\% of the total \( 2\, \mu \mathrm{m} \)
flux there. The fainter stars with the featureless reddened continua could be
main-sequence O stars. The four stars in the \( \SgrA  \) cluster (inner \( 1\as  \))
for which spectra were obtained appear to be such early-type stars (Genzel et
al. \cite{Genzel97}). Krabbe et al. (\cite{Krabbe95}) modeled the \HeI stars
and the late-type supergiants in the central parsec by a \( 3000\Mo  \) starburst
that occurred between \( 3 \) and \( 8\times 10^{6} \) yr ago. This is a very
small fraction of the total stellar mass that is enclosed in the central parsec,
\( \sim 1.5\times 10^{6}\Mo  \). These recently formed stars are not dynamically
relaxed, as is demonstrated by the radial velocity field of the \HeI stars in
the inner \( \sim 10\as  \), which clearly indicates that the stars are counter-rotating
at a constant rotation velocity \( v_{\mathrm{rot}}=120\, \mathrm{km}\, \mathrm{s}^{-1} \)
with respect to the galactic rotation of \( v_{\mathrm{rot}}=24\, \mathrm{km}\, \mathrm{s}^{-1} \).
There are additional indications that the stellar population varies with position
and is not quite well-mixed. The depth of the CO absorption feature in the integrated
light decreases in the inner \( \sim 10\as  \) (Sellgren et al. \cite{Sellgren90}).
This is probably due to the joint effects of a depletion of the luminous (\( K<10.5\kmag  \))
late-type stars in the inner \( 5\as  \) and the dilution of the absorption
feature by the concentration of luminous early-type stars there (Krabbe et al.
\cite{Krabbe95}). 

The basic premise of the attempts to measure the global properties of the stellar
distribution in the inner GC is that there is an underlying dynamically relaxed
distribution, which includes most of the mass and most of the stars, and which
is traced by the old faint stars. In contrast, the young stars and bright giants,
which dominate the light, are less reliable tracers because they could be unrelaxed,
or subject to statistical fluctuations due to their small numbers, or susceptible
to environmental effects such as stellar collisions or photoionization (Sellgren
et al. \cite{Sellgren90}). 

Our current understanding of the stellar distribution in the GC is incomplete.
The light distribution approximately follows a \( r^{-2} \) power-law from
as far away as \( \sim 100 \) pc, where it cannot be directly related to the
central super-massive BH, to \( \sim 1 \) pc from \( \SgrA  \) (Serabyn \&
Morris \cite{SM96}). Previous attempts to characterize the distribution in
the inner parsec concentrated on identifying the \emph{core radius}, \( r_{c} \)
, where the surface density falls to half its central value. The value of \( r_{c} \)
is much debated in the literature, and ranges from \( 0.05 \) pc (Allen \cite{Allen83})
to \( \sim 1 \) pc (Rieke \& Rieke \cite{Rieke88}). It appears that the differences
between the various estimates depend on whether the light distribution or the
number counts are used and whether or not the bright stars are included in the
fits. 

The most recent estimates for the mass distribution in the inner GC are dynamical
and are based on the stellar apparent motions and radial motions. Genzel et
al. (\cite{Genzel97}) and Ghez et al. (\cite{Ghez98}) fit the enclosed mass
estimates to a flattened isothermal distribution of the form

\begin{equation}
\label{eq:soft}
\rho (r)=\frac{\rho _{0}}{1+3(r/r_{c})^{2}}\, ,
\end{equation}
where \( r \) is the 3-dimensional distance from the BH, \( \rho _{0}=4\times 10^{6}\Mo \mathrm{pc}^{-3} \)
and \( r_{c}=0.4 \) pc. Because the BH mass completely dominates the velocity
field at \( r\ll r_{c} \), and the stellar mass begins to affect the stellar
velocities only at \( r\sim r_{c} \), this mass distribution measures only
the \emph{integrated} stellar mass up to \( r\sim r_{c} \), and is insensitive
to the distribution of the stars within the core volume. Therefore, the good
fit of the velocity field to this mass model does not necessarily preclude the
existence of a stellar cusp in the core.

\subsection{The \protect\( K\protect \) luminosity function and mass function}

\label{sec:KLF}

The construction of the \( K \) luminosity function (KLF) and mass function
models for the GC is discussed in detail by Alexander \& Sternberg (\cite{AS99}).
Briefly, the observed KLF in the GC (Blum et al. \cite{Blum96}; Davidge et
al. \cite{Davidge97}) can be smoothly joined to that in the Galactic Bulge,
as observed through Baade's window (Tiede et al. \cite{Tiede95}), and further
extended to lower luminosities by transforming deep optical counts in the bulge
(Holtzman et al. \cite{Holtzman98}) to the \( K \)-band. The resulting KLF
(fraction of stars per \( K \)-magnitude) has the form \( df/dK\propto 10^{bK} \)
with \( b=0.35 \), and extends from brightest observed stars in the GC at \( K_{1}=8^{\mathrm{m}} \)
down to the main-sequence turnoff at \( K_{2}\sim 21.5^{\mathrm{m}} \). Stars
below the main-sequence turnoff at \( \sim 1\, \Mo  \) have negligible contribution
to the KLF. By normalizing the KLF to the observed star counts in the GC, it
can be verified that it reproduces the observed mass-to-light ratio in the GC.
This empirical KLF and mass-to-light ratio can also be reproduced remarkably
well by a population synthesis model that assumes a continuous star-formation
history with a Miller-Scalo (Miller \& Scalo \cite{MS79}; Scalo \cite{Scalo86})
initial mass function (IMF) with a mass range of \( 0.1\Mo  \) to \( 120\Mo  \).
The IMF can be then used to follow the evolution of the stellar mass function
over time (\S\ref{sec:coll}).

We adopt these luminosity and mass functions as a volume averaged description
of the stellar population in the inner parsecs of the GC and assume that they
adequately describe the ``well-mixed'' population.

\subsection{The observed star counts }

\label{sec: obs}

\subsubsection{The Keck data set}

\label{sec:Keck}

\begin{figure}[tbp]
{\par\centering \resizebox*{!}{0.35\textheight}{\includegraphics{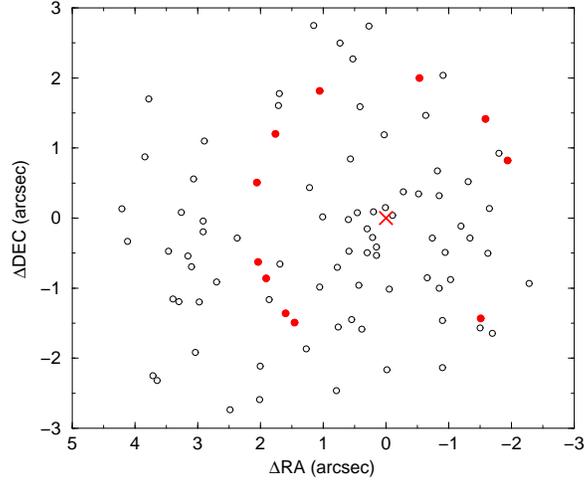}} \par}

\caption{\label{fig:ghez}90 stars in the inner \protect\( \sim 6''\times 6''\protect \)
observed by the Keck interferometer (Ghez et al. \cite{Ghez98}). Stars marked
by filled circles constitute the ``ring'' feature (Appendix \S\ref{sec: ring}).
Only the 64 stars in the central \protect\( 2.5\as \protect \) (of which 10
are early-type stars) are used in the analysis.}
\end{figure}

Ghez et al. (\cite{Ghez98}) obtained diffraction limited (\( 0.05\as  \) resolution)
\( K \)-band images of the inner \( 5''\times 5'' \) with the Keck interferometer,
down to a detection threshold of \( K_{0}=17\kmag  \). Using conservative selection
criteria, they extracted from the field 90 stars for which the angular motions
could be reliably measured and tabulated their magnitude and projected positions
and velocities (Figure \ref{fig:ghez}). As the field is slightly off-center
with respect to \( \SgrA  \), only the inner projected \( \sim 2.5\as  \)
are covered over the full \( 2\pi  \) radians. The original criterion for inclusion
in the data set was the reliability of measuring angular motion, rather than
completeness in magnitude. Consequently, the counts are complete only down to
\( \sim 14\kmag  \)--\( 15\kmag  \), beyond where the completeness falls off
very rapidly (A. Ghez, private comm.) to a level of \( \ga 1\% \) in the range
\( \sim 16\kmag  \)--\( 17\kmag  \). The incompleteness is not uniform over
the field, as it depends on the distribution of the bright stars. In our analysis
we consider only the 64 stars in the inner \( 2.5\as  \), of which 10 are identified
as early type stars. We assume here that the remaining 54 spectrally unidentified
stars are late type stars.

\subsubsection{The OSIRIS data set }

\label{sec:Osiris}

\begin{figure}[tbp]
{\par\centering \resizebox*{!}{0.35\textheight}{\includegraphics{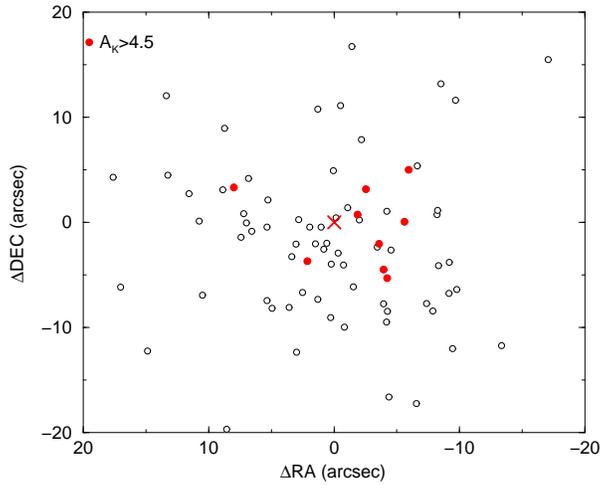}} \par}

\caption{\label{fig:blum}The inner \protect\( 40''\times 40''\protect \) of the OSIRIS
field (Blum et al. \cite{Blum96}). Stars marked by filled circles are highly
reddened stars (\protect\( A_{K}>4.5^{\mathrm{m}}\protect \)). Only the 50
late-type stars in the inner \protect\( 13\as \protect \) are used in the analysis. }
\end{figure}

Blum et al. (\cite{Blum96}) imaged the inner \( 2' \) of the GC with the Ohio
State Infrared Imager and Spectrometer (OSIRIS) in the \( J \), \( H \) and
\( K \) bands (\( \sim 1\as  \) resolution) for the purpose of measuring the
luminosity function and color-magnitude diagram of the stellar population. They
detected a total of \( \sim 2000 \) stars and determined by simulations that
the entire sample is complete down to \( K\sim 12\kmag  \). Of this large sample,
they tabulated the \( K \) magnitude, projected positions and \( K \) extinction
coefficient \( A_{K} \) for a sub-sample of 147 stars, which span a range of
\( K=6.4\kmag  \)--\( 12.4\kmag  \). The stars included in the sub-sample
are those brighter than \( K=10.5\kmag  \), those for which Blum et al. obtained
\( 2\mu \mathrm{m} \) spectra, or those with IRS numbers. Thus, this sub-sample
is guaranteed to be complete only down to \( K=10.5\kmag  \). The distribution
of the stars in the field is very sparse beyond projected radius \( p\sim 10\as  \)--\( 15\as  \)
(Figure~\ref{fig:blum}). We limit our analysis to the 50 late type stars in
the inner \( 13\as  \) in order not to be affected too much by the Poisson
fluctuation in the counts or by molecular gas and dust that may extend inward
from the circumnuclear ring (Genzel, Hollenbach \& Townes \cite{GHT94}).

\subsubsection{The 3D and SHARP data set }

\label{sec:3D+SHARP}

Genzel et al. (\cite{Genzel96}) used the 3D imaging spectrometer to identify
spectrally 223 stars in the central parsec and separate them into early and
late-type stars. The stars were assigned to approximate magnitude classes: bright
(\( K\le 10.5\kmag  \)), intermediate (\( 10.5\kmag <K\le 12\kmag  \)) and
faint (\( 12\kmag <K \)). Because the 3D counts are incomplete in the inner
\( 5\as  \) due to source confusion, they augmented the counts with high resolution
SHARP camera observations of \( K<13.5^{\mathrm{m}} \) stars in the inner \( 10\as  \)
(Eckart \& Genzel \cite{EG97}), but did not include the 11 very faint (\( K>14^{\mathrm{m}} \))
\( \SgrA  \) cluster stars (\( S1 \)-\( S11 \)) in the inner \( \sim 0.5\as  \)
(A. Eckart, private comm.). With the faintest stars excluded from the sample,
the composite data set is of roughly uniform depth over the central \( \sim 20\as  \).

\begin{figure}[tbp]
{\par\centering \resizebox*{!}{0.35\textheight}{\includegraphics{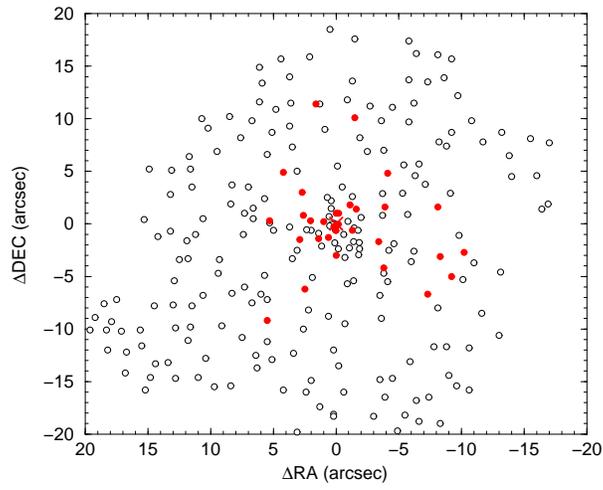}} \par}

\caption{\label{fig:genzel}The combined 3D (Genzel et al. \cite{Genzel96}) and SHARP
(Eckart \& Genzel \cite{EG97}) counts in the inner \protect\( 20\as \protect \).
Stars marked by filled circles are identified \HeI and early-type stars. Only
the 92 late-type 3D stars and the 15 spectrally unidentified SHARP stars (excluding
the faint \protect\( \SgrA \protect \) cluster stars) in the inner \protect\( 13\as \protect \)
are used in the analysis.}
\end{figure}

Genzel et al. (\cite{Genzel96}) binned the composite data set and fitted a
flattened \( r^{-1.8} \) power-law distribution (analogous to Eq.~\ref{eq:soft})
to the surface number density and obtained a core radius of \( r_{c}=0.29_{-0.09}^{+0.19} \)
pc (\( r_{c}=7.4\as \, _{-1.6}^{+5.0} \))\footnote{
This is a somewhat smaller value than \( r_{c}=10\as  \) that was adopted in
their subsequent works as a compromise between the different core radii suggested
in the literature (R. Genzel, private comm.). 
}. We reconstruct here the inner \( \sim 20\as  \) of the the unbinned composite
data set for from the tabulated 3D and SHARP stellar positions in order to check
whether other density models can fit these star counts as well. Our composite
data set includes in total 213 stars, of which 198 are late-type 3D stars from
Genzel et al. (\cite{Genzel96}) and 15 are spectrally unidentified SHARP stars
from Eckart \& Genzel (\cite{EG97}). The surface distribution of the 213 stars
is shown in Fig.~\ref{fig:genzel}. To avoid edge-of-field effects, we consider
in our analysis below only the 107 late-type stars in the inner \( 13\as  \),
which are composed of 92 3D late-type stars and the 15 SHARP stars.

\section{Statistical analysis of the star counts}

\label{sec:stat}

\subsection{Stellar density distribution models \label{sec:theory}}

The main purpose of this work is to quantify the empirical stellar density distribution
in the GC, but theoretical expectations guide us in the choice of models for
describing the distribution. Generally, the effect of a super-massive BH on
the stellar population within the cusp radius will depend on whether the relaxation
time due to two-body stellar encounters is longer or shorter than the system's
age. When two-body relaxation is unimportant, the black hole may grow adiabatically,
and the resulting stellar distribution will depend on the initial conditions.
A spherically symmetric, non-rotating system that is initially isothermal will
develop a \( r^{-3/2} \) cusp (Young \cite{Young80}). Other initial conditions
will result in a variety of density distributions, in some cases as steep as
\( r^{-5/2} \) (Lee \& Goodman \cite{LG89}; Quinlan, Hernquist \& Sigurdsson
\cite{Quinlan95}). 

When two-body encounters are effective, the final configuration will not depend
on the initial one. Very close to the BH, where inelastic stellar collisions
dominate, a \( r^{-1/2} \) cusp of marginally bound stars will form (Murphy,
Cohn \& Durisen \cite{MCD91}). At larger radii, where collisions are elastic,
a single mass population will settle into a \( r^{-7/4} \) cusp (Bahcall \&
Wolf \cite{BW76}). In a realistic stellar population with a spectrum of masses,
mass segregation will take place, and the massive stars will be more concentrated
towards the center than the low-mass stars. Bahcall \& Wolf (\cite{BW77}) investigated
the two-body relaxation of a two-component stellar population around a massive
BH. They found that the two components settle into different power-law distribution
with power-law indices given by
\begin{equation}
\label{eq:mslope}
\alpha _{i}=\frac{m_{i}}{4m_{2}}+\frac{3}{2}\qquad (m_{1}<m_{2}).
\end{equation}
 Murphy, Cohn \& Durisen (\cite{MCD91}) showed that Eq.~\ref{eq:mslope} holds
also for the spatial distribution of a multi-mass population, when \( m_{2} \)
is taken to be the maximal mass cutoff, so that \( 3/2\le \alpha _{i}\le 7/4 \).
It is straightforward to verify that over a limited range in \( r \), the total
density distribution that will be observed will still be an approximate power-law
with an index in this range. 

The local two-body relaxation time at the core radius of the dynamical mass
model (Eq.~\ref{eq:soft}) can be estimated by (Spitzer \& Hart \cite{SH71})
\begin{equation}
\label{eq:trelax}
t_{r}\approx \frac{0.34\sigma ^{3}}{G^{2}\left\langle m\right\rangle \rho \ln \Lambda }\sim 3\, 10^{9}\, \mathrm{yr}\, ,
\end{equation}
 where \( \Lambda \approx M_{\bullet }/\left\langle m\right\rangle  \) and
we assume \( \left\langle m\right\rangle =0.6\, \Mo  \), a BH mass of \( M_{\bullet }=2.6\times 10^{6}\, \Mo  \)
(Genzel et al. \cite{Genzel97}; Ghez et al. \cite{Ghez98}) and \( \rho =10^{6}\, \Mo \mathrm{pc}^{-3} \),
\( \sigma =117\, \mathrm{km}\, \mathrm{s}^{-1} \)at \( r_{c}=0.4 \) pc (Genzel
et al. \cite{Genzel96}). Since the relaxation time is shorter than the age
of the Galaxy, the theoretical expectation is that the observed giants, which
have an intermediate mass, will have a power-law distribution with an index
in the range \( 3/2<\alpha <7/4 \). However, we will avoid here any theoretical
prejudice and investigate power-law distributions with indices that span the
full range of the various theoretical scenarios from \( \alpha =0 \) to \( 5/2 \).

We consider two \emph{}spherically symmetric stellar density models. The first
is the flattened isothermal model (Eq.~\ref{eq:soft}), where we assume a constant
ratio between the mass density and the stellar number density

\begin{equation}
\label{eq:rhoiso}
n(r)=\frac{n_{0}}{1+3(r/r_{c})^{2}},
\end{equation}
 and where \( n_{0} \) is the central number density and \( r_{c}=10'' \)
(Genzel et al. \cite{Genzel96}). The corresponding surface density is

\begin{equation}
\label{eq:sigsio}
\Sigma (p)=\frac{\pi r_{c}^{2}n_{0}}{3\sqrt{p^{2}+r_{c}^{2}\left/ 3\right. }},
\end{equation}
 where \( p \) is the projected distance from the center. The second class
of models are broken power-laws,

\begin{equation}
\label{eq:rhopl}
n(r)=\left\{ \begin{array}{lr}
n_{b}\left( r/r_{b}\right) ^{-\alpha } & r_{0}<r<r_{b}\\
n_{b}\left( r/r_{b}\right) ^{-\beta } & \qquad r\geq r_{b},\, \beta >1
\end{array}\right. ,
\end{equation}
 where \( n_{b} \) is the density at the break at \( r_{b} \) , and \( r_{0} \)
is the closest distance to the BH where stars can exist. Although \( r_{b} \)
is \emph{not} the core radius, we will loosely refer to the region interior
to \( r_{b} \) as the core. The corresponding surface density is

\begin{equation}
\label{eq:sigpl}
\sum (p)=n_{b}\left\{ \begin{array}{lr}
r_{b}^{\beta }B\left( \frac{1}{2},\frac{\beta -1}{2}\right) p^{1-\beta }+2\sqrt{r_{b}^{2}-p^{2}} & \\
\, \, \, \, \, \, \times \left( H\left( \alpha ,\frac{p}{r_{b}}\right) \left( \frac{p}{r_{b}}\right) ^{-\alpha }-H\left( \beta ,\frac{p}{r_{b}}\right) \left( \frac{p}{r_{b}}\right) ^{-\beta }\right) \, \, \,  & r_{0}<p<r_{b}\\
r_{b}^{\beta }B\left( \frac{1}{2},\frac{\beta -1}{2}\right) p^{1-\beta } & p>r_{b}
\end{array}\right. 
\end{equation}
 where \textbf{\( B \)} is the beta function and \( H \) is defined by the
hypergeometric function

\begin{equation}
H(x,y)=_{2\! \! \! }F_{1}\left( \frac{1}{2},\frac{x}{2},\frac{3}{2},1-y^{-2}\right) \, .
\end{equation}

\subsection{The cumulative distribution function}

\label{sec:KS}

\begin{figure}[tbp]
{\centering \begin{tabular}{cc}
\resizebox*{!}{0.3\textheight}{\includegraphics{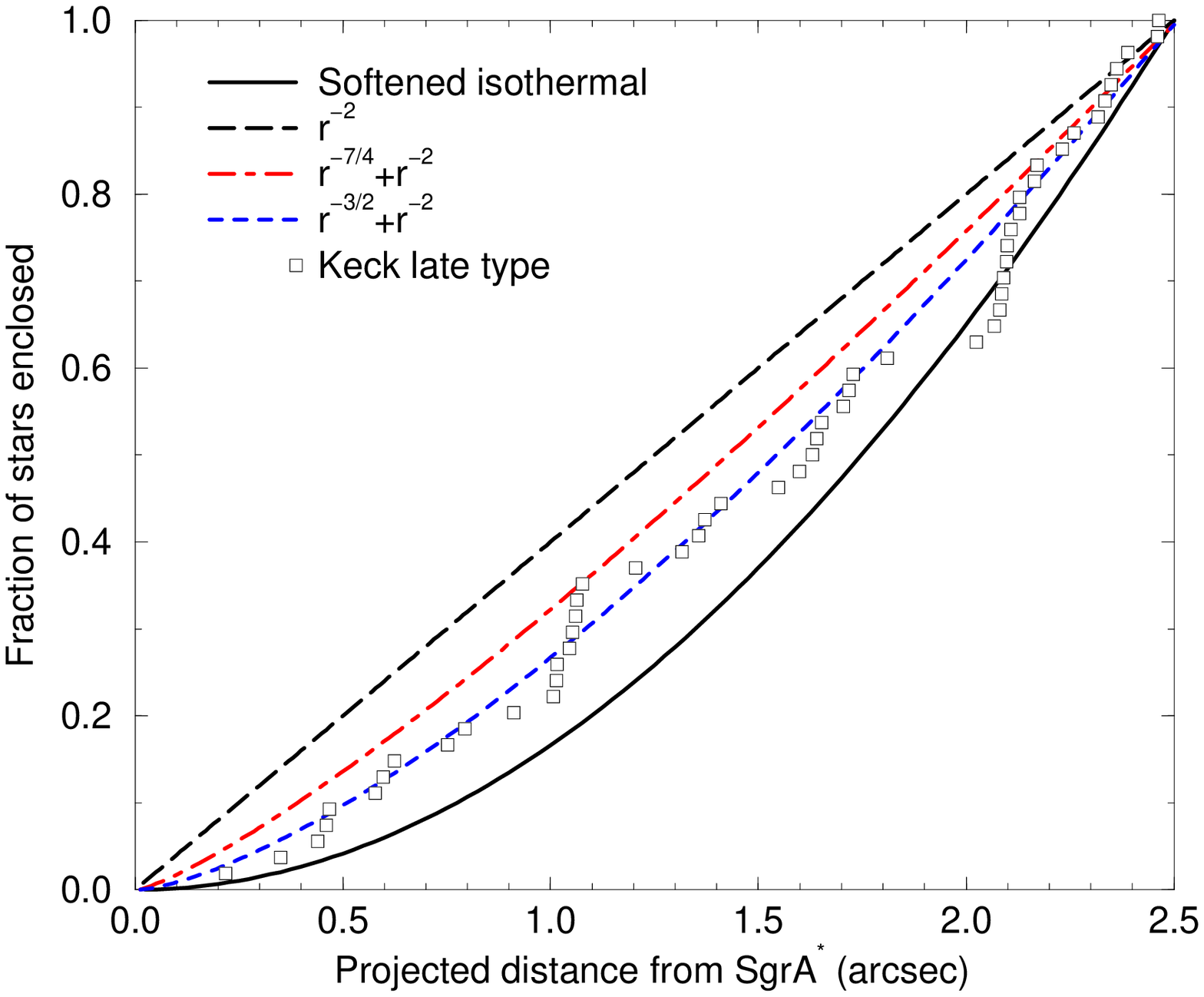}} &
\resizebox*{!}{0.3\textheight}{\includegraphics{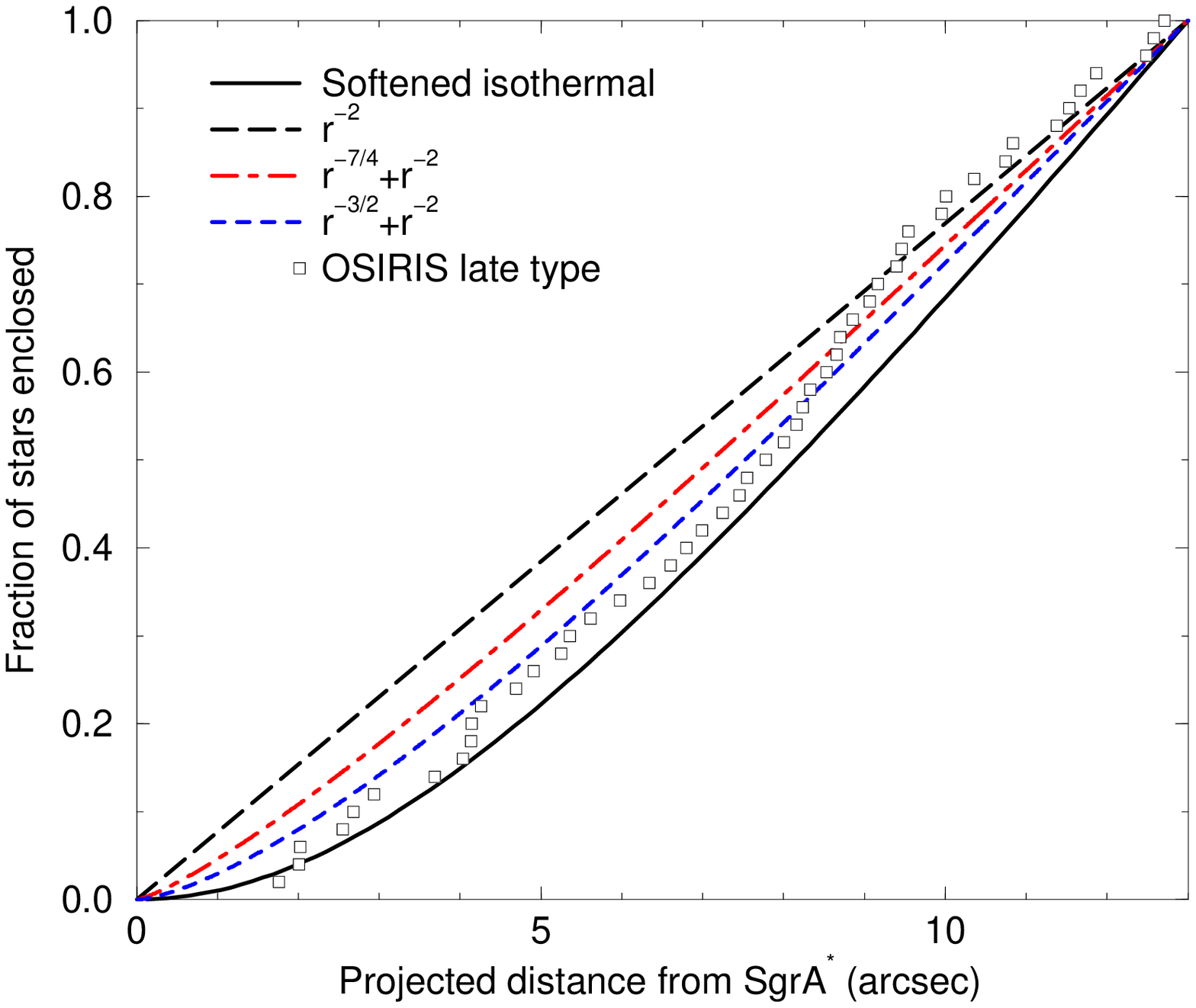}} \\
\multicolumn{2}{c}{\resizebox*{!}{0.3\textheight}{\includegraphics{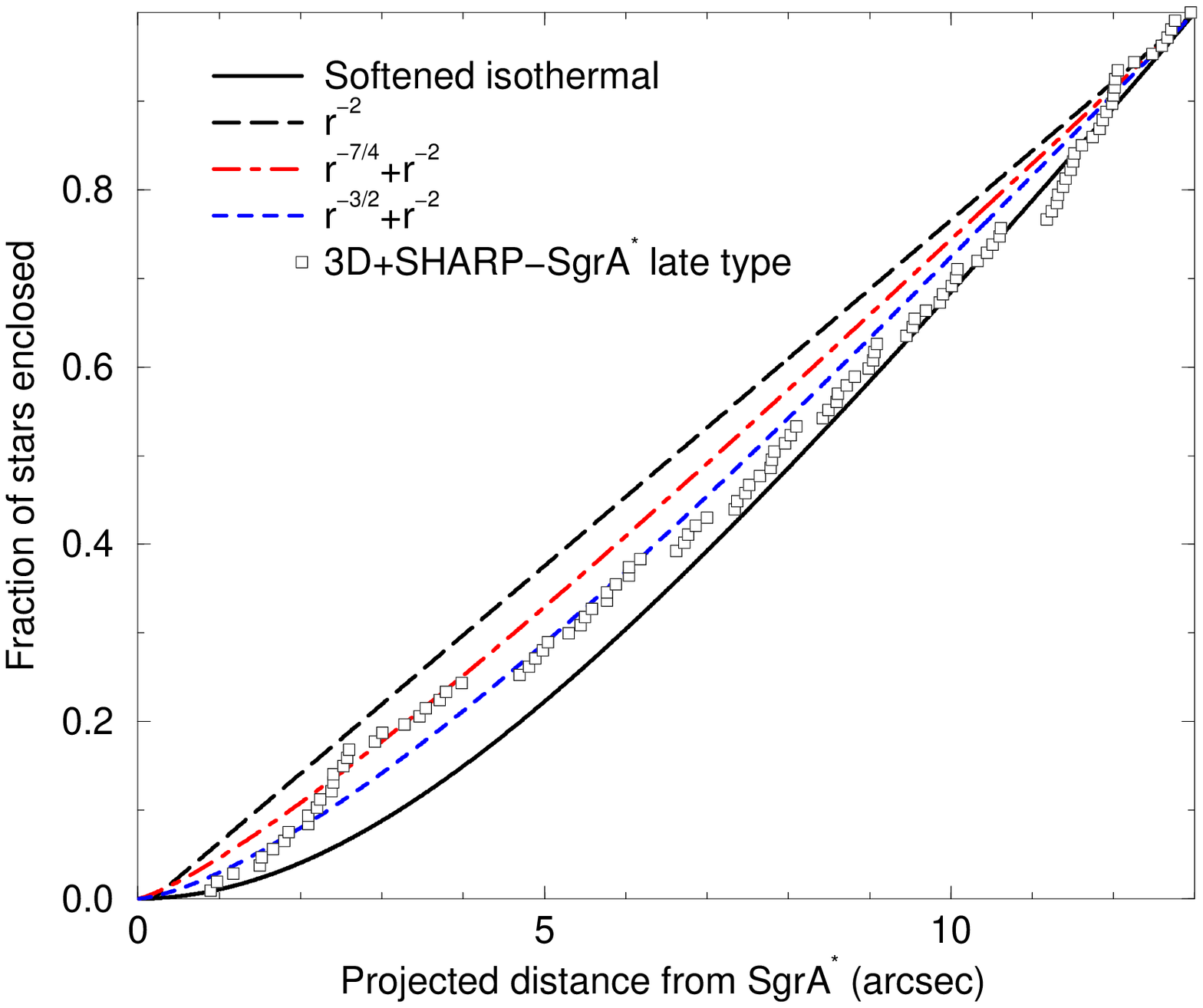}} }\\
\end{tabular}\par}

\caption{\label{fig:cum}The cumulative DF of the late-type stars for the different
data sets, compared to various stellar density distribution models. Top left:
the inner \protect\( 2.5''\protect \) of the Keck field (Ghez et al. \cite{Ghez98}).
Top right: the inner \protect\( 13\as \protect \) of the OSIRIS field (Blum
et al. \cite{Blum96}). Bottom: the inner \protect\( 13\as \protect \) of the
3D field (Genzel et al. \cite{Genzel96}) together with the inner \protect\( 10\as \protect \)
of the SHARP data excluding the faint \protect\( \SgrA \protect \) cluster
stars (Eckart \& Genzel \cite{EG97}).}
\end{figure}

Figure~\ref{fig:cum} shows the cumulative distribution functions (DFs) for
the late-type stars in the three data sets (more precisely, all the observed
stars that were not positively identified as \HeI stars, early-type stars or
dust embedded stars). The cumulative DFs indicate that neither a flattened isothermal
model (Eq.~\ref{eq:rhoiso}) nor a singular isothermal model \( n\propto r^{-2} \)
provide a good description of the observed star counts in the central \( \sim 10'' \).
This impression is supported by the Kolmogorov-Smirnov (K-S) test results (presented
as acceptance probabilities in Table~\ref{tab: ks}), although the K-S scores
do not differentiate strongly between the different models. Generally, the K-S
scores favor cusps with slopes in the range \( \alpha \sim 3/2 \) to \( 7/4 \).
The discrepancy between the OSIRIS data and the flattened isothermal model is
the smallest of all data sets, although this is not the best-fitting model.
This probably reflects the fact that the OSIRIS data set is less deep, and is
therefore more sensitive to the decrease in the number of luminous late-type
giants in the central few arcseconds.

The statistical significance of the concentration of stars at \( p\sim 2'' \),
which is seen in the cumulative distribution of the Keck data set, is discussed
in appendix~\ref{sec: ring}.

\subsection{Maximum Likelihood analysis}

\label{sec:ML}

\begin{figure}[tbp]
{\centering \begin{tabular}{cc}
\resizebox*{!}{0.3\textheight}{\includegraphics{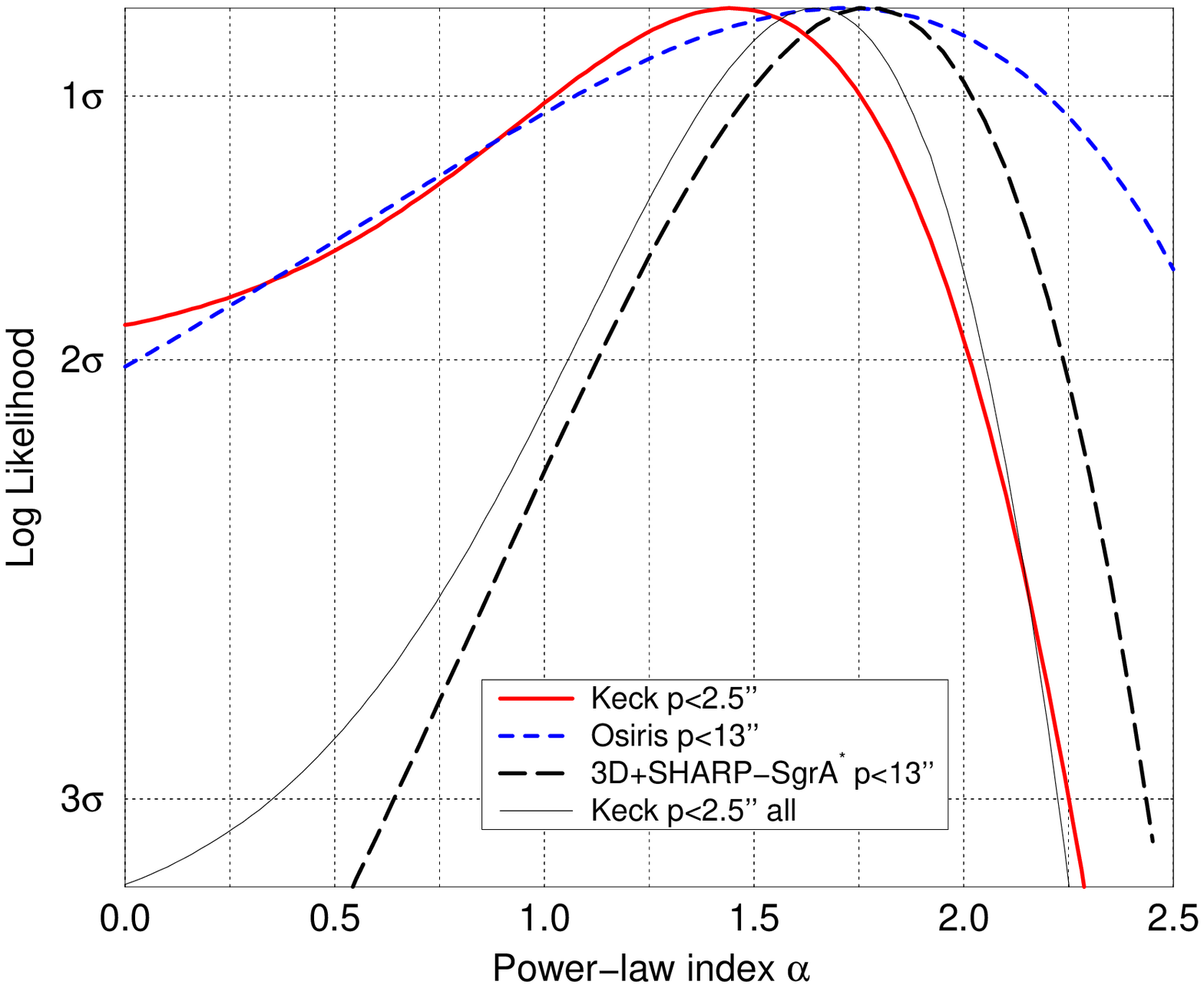}} &
\resizebox*{!}{0.3\textheight}{\includegraphics{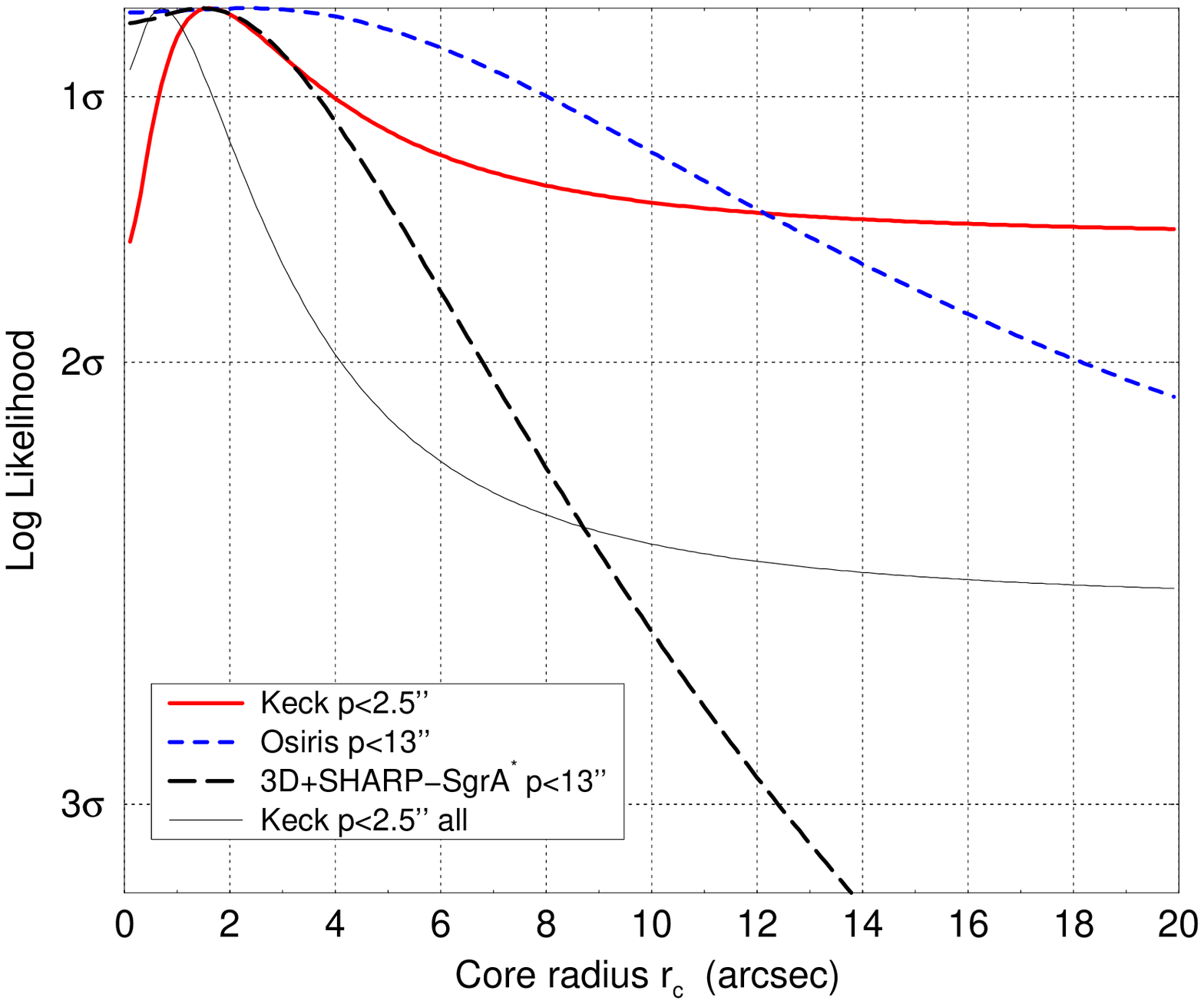}} \\
\end{tabular}\par}

\caption{\label{fig:ML}The ML estimate of the stellar distribution for the late-type
stars in the three data sets. The likelihood curves for the full Keck data set
(including early-type stars) is also presented for comparison. Left: The likelihood
curves for the power-law index \protect\( \alpha \protect \) of the broken
power-law distribution, for fixed values of \protect\( \beta =2\protect \),
\protect\( r_{b}=10\as \protect \). Right: The likelihood curves for the core
radius \protect\( r_{c}\protect \) of the flattened isothermal distribution.}
\end{figure}

The trends that are seen in the cumulative distributions of the star counts
(Figure~\ref{fig:cum}) and the theoretical expectations (\S\ref{sec:theory})
suggest fitting power-law density distributions to the star counts. In order
to avoid the arbitrariness of binning, we perform a Maximum Likelihood (ML)
analysis to find the best fitting power-law index \( \alpha  \), assuming a
broken power-law distribution with fixed values \( \beta =2 \) and \( r_{b}=10\as  \)
(we verified that the results are not very sensitive to the exact shape of the
outer distribution). Figure~\ref{fig:ML} shows the likelihood curves that were
obtained by using Eq.~\ref{eq:sigpl} to calculate the relative probabilities
of the observed stellar positions\footnote{
For each density model, we normalize the probability to the projected distance
interval between the innermost and outermost observed stars in the sample. This
is necessary for removing biases due to the finite size of the field, and biases
due to divergent behavior of the density at \( p=0 \), where the measured positions
are not much larger than the angular resolution and the \( \sim 0.1\as  \)
uncertainty in the position of the dynamic center (Ghez et al \cite{Ghez98}). 
}. As can be expected, the ML best-fit parameters are not exactly the same as
those derived from the Kolmogorov-Smirnov analysis, since the two methods are
sensitive to different ranges of the data (the K-S to the mid-range of the projected
distance and the ML to the inner, high-probability region). Nevertheless, the
results are qualitatively similar. The likelihood curves for all the samples
peak in the range \( \alpha \sim 1.5 \) to \( 1.75 \), irrespective of whether
the they are based on stars in the inner \( 13\as  \), or only the inner \( 2.5\as  \).
It is reassuring that this range is consistent with the ML estimate for the
entire Keck data set, as this suggests that the contamination by the \( \sim 20\% \)
of the young stars does not introduce a large bias. A flat core (\( \alpha =0 \))
is ruled out only at at the \( 2\sigma  \) level by the Keck and OSIRIS data
sets, but is completely incompatible with the 3D+SHARP data. 

We also performed a ML analysis to find the best fitting core radius \( r_{c} \),
assuming a flattened isothermal distribution. The results are qualitatively
similar to those of the broken power-law. In all data sets the likelihood curves
peak at \( r_{c}<2.5\as  \), and while a \( 10\as  \) core radius is still
compatible with the Keck and OSIRIS data sets at the \( \sim 1.5\sigma  \)
level, it is ruled out by the 3D+SHARP data at the \( \sim 2.5\sigma  \) level. 

The best fit core radius we derive from our composite 3D+SHARP data set for
the late-type stars (\( r_{c}=1.5\as  \)) is much smaller than the one derived
by Genzel et al. (\cite{Genzel96}) (\( r_{c}=7.4\as \, _{-1.6}^{+5.0} \)).
The reasons for this disparity are unclear. The most significant difference
between the two analysis methods appears to be the issue of binning. The binned
star counts that were used in the \( \chi ^{2} \) fit of Genzel et al. (\cite{Genzel96})
could lead to the suppression of a steep cusp in sparse counts. Our initial
attempts to fit density models to binned 3D+SHARP data indicated that the best-fit
density in the innermost regions fluctuates considerably with small changes
in the bin sizes or bin centers. A more detailed comparison of the unbinned
ML and the binned \( \chi ^{2} \) fitting procedures, which is outside the
scope of this work, will be needed to resolve this issue.

\section{Extinction in the inner GC}

\label{sec:Ak}

A gradient in the extinction field could bias the star-counts because fewer
stars would be observed where the extinction is higher. For example, a central
concentration of dust could make a stellar cusp appear as a flat core, and conversely,
a dust ring could make a flat core appear as a cusp. In this section we use
observations of the extinction in the GC to estimate the magnitude of such a
bias.

\begin{figure}[tbp]
{\par\centering \resizebox*{!}{0.35\textheight}{\includegraphics{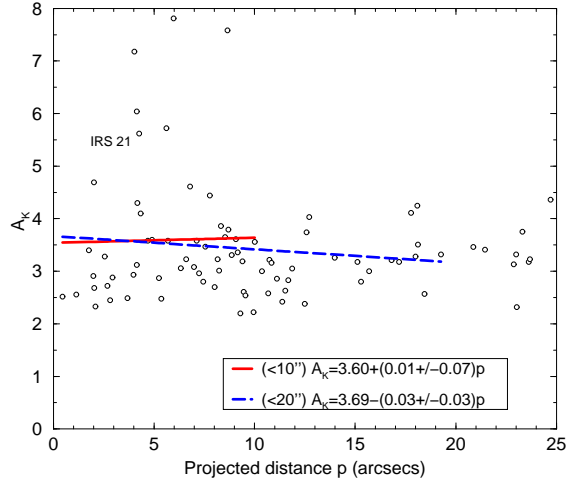}} \par}

\caption{\label{fig:AK}The extinction of individual stars in the inner \protect\( 25\as \protect \)
of the GC deduced from the near infrared colors (Blum et al. \cite{Blum96}).
The average error on the extinction estimate is \protect\( \sim 0.1\kmag \protect \)
(not including systematic error due to the assumed intrinsic colors). The linear
regression of \protect\( A_{K}\protect \) on \protect\( p\protect \) in the
inner \protect\( 10\as \protect \) and \protect\( 20\as \protect \) is also
shown. IRS 21 is a highly polarized extended object (Ott, Eckart \& Genzel \cite{Ott98}). }
\end{figure}

Davidge (\cite{Davidge98}) used \( J \), \( H \), and \( K \) observations
of stars in the inner \( 3' \) to estimate the radial dependence of the \( K \)-band
extinction coefficient, \( A_{K} \), averaged over \( 20\as  \)-wide bins.
Davidge found that \( \left\langle A_{K}\right\rangle  \) systematically increases
from \( 2.8^{\mathrm{m}} \) at \( \sim 100\as  \) to \( 3.2^{\mathrm{m}} \)
at \( \sim 30\as  \), and then decreases again at the inner \( 20'' \) to
\( 2.9^{\mathrm{m}} \) (\( \Delta A_{K}\sim -0.02^{\mathrm{m}} \) per arcsec).
Blum et al. (\cite{Blum96}) made a similar study of the inner \( 2' \), and
obtained an average central extinction value of \( \left\langle A_{K}\right\rangle \sim 3.3^{\mathrm{m}} \).
Eckart et al. (\cite{Eckart95}) found no significant color gradient over the
\( \SgrA  \) cluster in the inner \( 1\as  \). Figure~\ref{fig:AK} shows
the distribution of stars in the \( A_{K}-p \) plane in the inner \( 25'' \)
of the GC (Blum et al. \cite{Blum96}). A small fraction of the stars, all concentrated
in the inner \( 10'' \), are heavily reddened. One of those (IRS 21) was identified
by Ott, Eckart \& Genzel (\cite{Ott98}) as an extended and highly polarized
object, which may be a recently formed star still embedded in a dust shell.
A second such star (IRS 1W) was also identified in the inner \( 10\as  \),
but is not part of the OSIRIS data set. 

\begin{figure}[tbp]
{\par\centering \resizebox*{!}{0.3\textheight}{\includegraphics{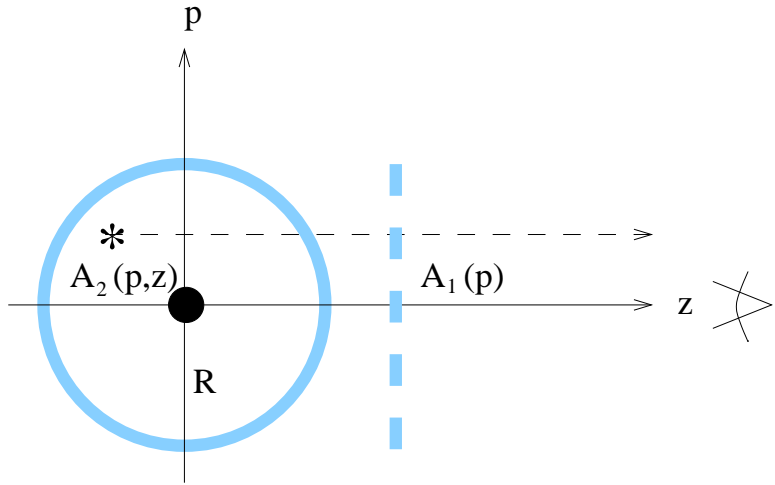}} \par}

\caption{\label{fig:dust}A simple trial model for the extinction field in the GC. A
star at projected distance \protect\( p\protect \) and point \protect\( z\protect \)
along the line-of-sight, is embedded in a spherical dust cloud of core radius
\protect\( R\protect \) that surrounds the black hole. The total extinction
is the sum of the extinction due to dust inside the cloud, \protect\( A_{2}(p,z)\protect \),
and to dust in a foreground screen, \protect\( A_{1}(p)\protect \).}
\end{figure}

\begin{figure}[tbp]
{\centering \begin{tabular}{c}
\resizebox*{!}{0.4\textheight}{\includegraphics{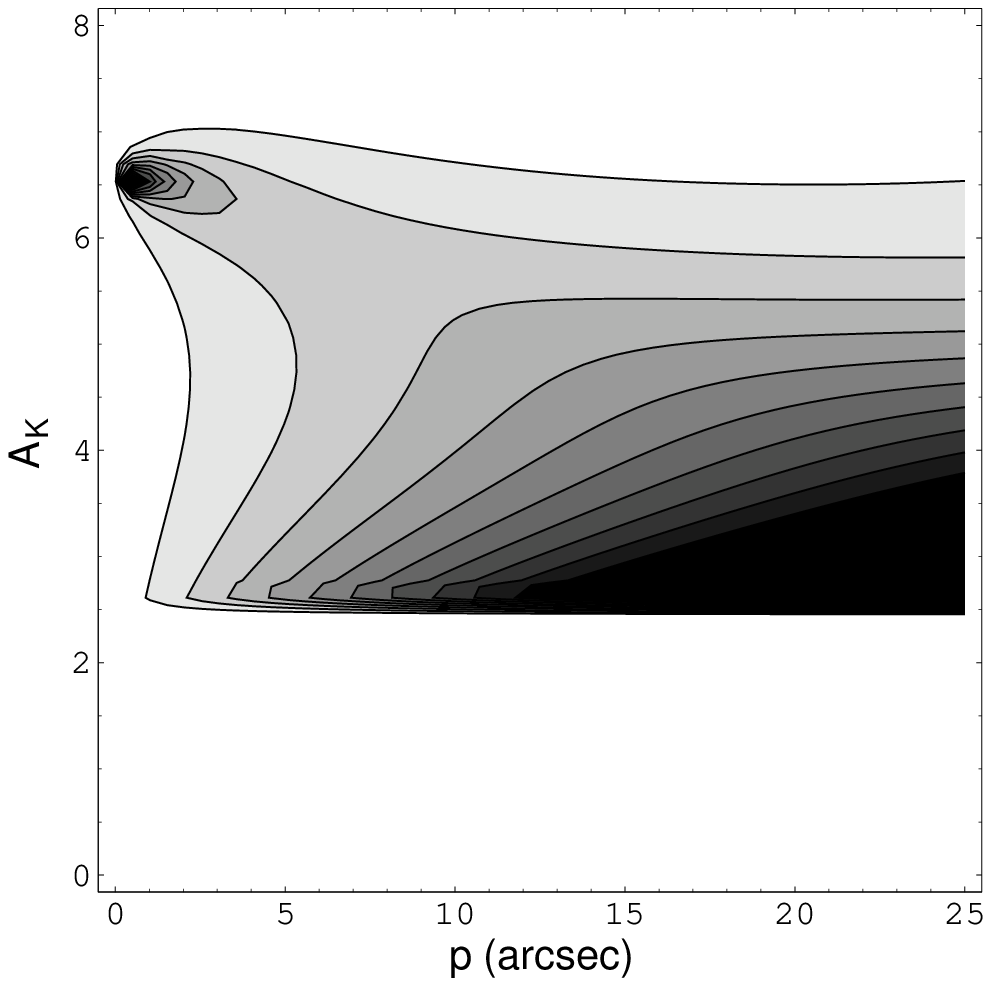}} \\
\resizebox*{!}{0.4\textheight}{\includegraphics{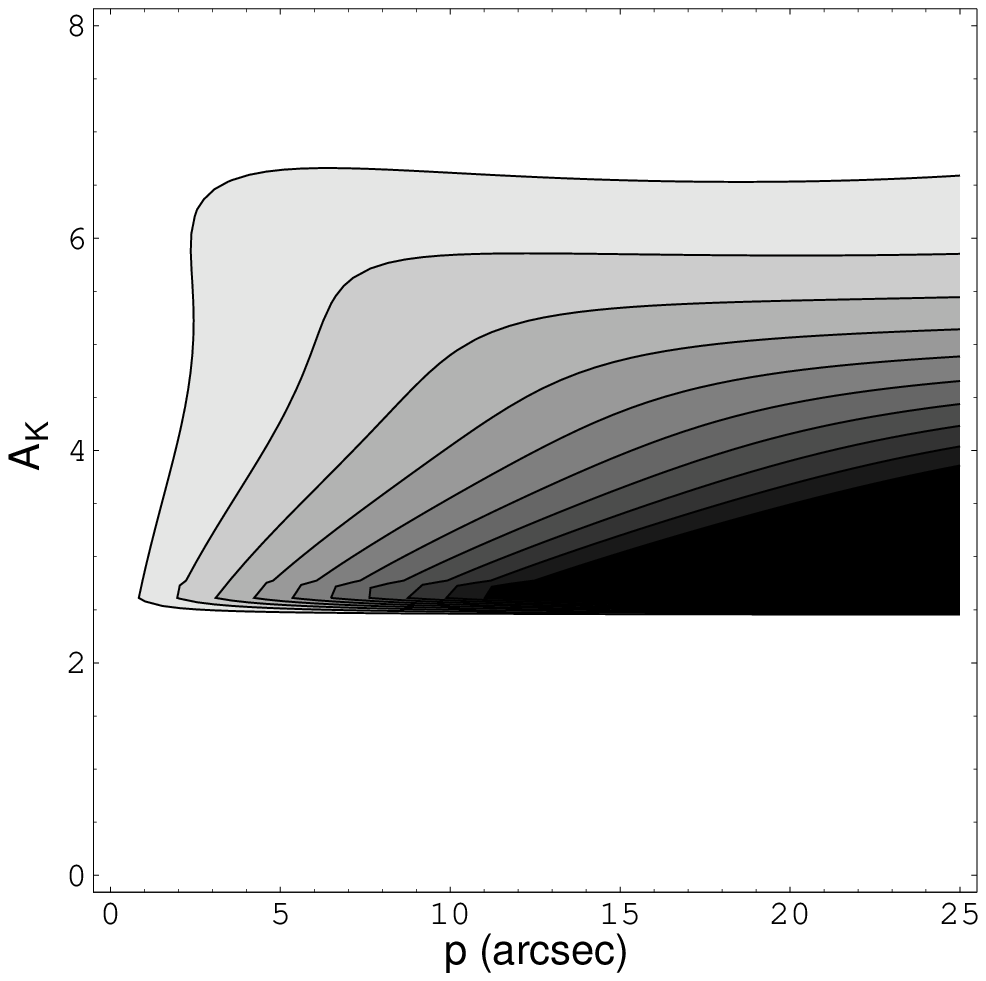}} \\
\end{tabular}\par}

\caption{\label{fig:pak}The relative probability density (normalized to the peak) for
the detection of stars at projected distance \protect\( p\protect \) and extinction
\protect\( A_{K}\protect \). The dark regions correspond to high detection
probability. Top: a \protect\( n\propto r^{-2}\protect \) stellar distribution.
Bottom: a flattened isothermal model (Eq.~\ref{eq:rhoiso}). The assumed dust
model is a spherical \protect\( d=1\left/ \left( 1+3(r/R)^{2}\right) \right. \protect \)
distribution with \protect\( R=30''\protect \), \protect\( D=0.15\protect \),
and \protect\( A_{1}=2.5^{\mathrm{m}}\protect \), the luminosity function is
\protect\( df/dK\propto 10^{bK}\protect \) with \protect\( b=0.35\protect \),
and the detection threshold is \protect\( K_{0}=13^{\mathrm{m}}\protect \).}
\end{figure}

The observed total surface density, \( \Sigma _{0} \), and the observed differential
surface density, \( d\Sigma _{0}/dA_{K} \), reflect the 3-dimensional distribution
of the dust. The \( K \)-extinction suffered by a star at projected distance
\( p \) and at point \( z \) along the line-of-sight can be decomposed into
two parts. The first term, \( A_{1}(p) \), corresponds to extinction by a foreground
dust screen, and the second term, \( A_{2}(p,z) \), corresponds to that by
dust inside the GC (Figure~\ref{fig:dust}). The total extinction, \( A_{K} \),
is then
\begin{equation}
\label{eq:AKpz}
A_{K}(p,z)=A_{1}(p)+A_{2}(p,z)=A_{1}(p)+D\int ^{\infty }_{z}d(p,z')dz'\, ,
\end{equation}
where \( d \) is the dust density and \( D \) is a factor that translates
the dust column to extinction. Assuming that the unreddened \( K \) luminosity
function (KLF) is constant over the volume of interest and assuming cylindrical
symmetry around the line of sight, the distribution of stars in the \( p-A_{K} \)
plane is given by \( 2\pi p(d\Sigma _{0}/dA_{K}) \), where

\begin{equation}
\label{eq:pak}
\frac{d\Sigma _{0}}{dA_{K}}(p;K_{0})=\left( \frac{\partial z}{\partial A_{K}}\right) _{p}n\left[ p,z(A_{K})\right] \int ^{K_{0}-A_{K}}_{K_{1}}\frac{df}{dK}dK\qquad \mathrm{for}\: A_{K}>A_{1}(p),
\end{equation}
and \( K_{0} \) is the detection limit. The observed total surface density
\( \Sigma _{0} \) will be different from the true one since

\begin{equation}
\label{eq:S0}
\Sigma _{0}(p;K_{0})=\int n(p,z)dz\int ^{K_{0}-A_{K}(p,z)}_{K_{1}}\frac{df}{dK}dK\, .
\end{equation}
 It is important to emphasize that the measured quantity \( \left\langle A_{K}(p)\right\rangle  \),
which is averaged over the \emph{observed} stars, is \emph{not} directly related
to the average expressed by Eq.~\ref{eq:S0}, since regions of high extinction
will be under-represented in the counts, and since intrinsically very luminous
stars, which can be observed even through high extinction, are very rare. Therefore,
the relatively small measured value \( \left\langle A_{K}\right\rangle \sim 3.5 \)
does not necessarily preclude the possibility that the innermost GC is hidden
by strong extinction. In order to test this, it is necessary to consider the
3-dimensional distribution of the dust. Because it is unclear what completeness
corrections have to be applied to the data presented in Figure~\ref{fig:pak},
we will restrict ourselves here to qualitative comparisons of some simple dust
geometries with the observed distribution of stars in the \( p-A_{K} \) plane. 

We consider first the case where the dust is concentrated in a foreground screen,
so that \( A_{K}(p,z)=A_{1}(p) \). If \( A_{1} \) is a constant, or varies
randomly over small enough angular scales, \( \Sigma _{0} \) will simply be
proportional to \( \Sigma  \), at least on average, and the extinction will
not bias the observed density distribution. This requires us to interpret the
few observed highly reddened stars as \emph{intrinsically} reddened objects
that do not trace an extended dust distribution. A linear regression of \( A_{K} \)
on \( p \) (Figure~\ref{fig:AK}) does not show any strong systematic trends
in the radial behavior of \( A_{K} \) in the inner \( 20'' \) and is thus
consistent with a foreground dust screen, where the scatter is due to random
fluctuations in the extinction, or measurement scatter, or both. The differential
extinction is \( \Delta A_{K}=+0.01^{\mathrm{m}} \) per arcsec based on the
the inner \( 10'' \), and \( \Delta A_{K}=-0.03^{\mathrm{m}} \) per arcsec
based on the inner \( 20'' \), and in both cases \( \Delta A_{K}=0 \) is within
the \( 1\sigma  \) errors. 

The bias due to a non-zero trend of this order will not be large. If a power-law
density distribution is estimated from the counts at points \( p_{1} \) and
\( p_{2} \), then the difference between the measured power-law index and the
true one is

\begin{equation}
\label{eq:ar}
\Delta \alpha =\frac{b(p_{2}-p_{1})\Delta A_{K}}{\log _{10}(p_{2}/p_{1})}\, ,
\end{equation}
which is only \( \Delta \alpha <0.1 \) for \( \Delta A_{K}=-0.03^{\mathrm{m}} \)
per arcsec and \( p_{1}=1\as  \), \( p_{2}=10'' \). 

Next, we consider the general case, where in addition to a foreground dust screen
the BH and the stars near it are embedded in a dust cloud. In particular, we
wish to address the possibility that the highly reddened stars in the central
\( 10'' \) are those few stars that are luminous enough to be observed even
from inside the dust cloud. We assume a simple configuration where \( A_{1} \)
is constant and the dust is distributed in a spherical cloud with a density
profile
\begin{equation}
d=\frac{1}{1+3(r/R)^{2}}\, ,
\end{equation}
where \( R \) is the core radius of the dust cloud (Figure~\ref{fig:dust}).
Figure~\ref{fig:pak} shows the relative probability density for observing stars
in the \( p-A_{K} \) plane, as calculated from Eqs.~\ref{eq:AKpz} and \ref{eq:pak},
for a \( n\propto r^{-2} \) stellar distribution and a flattened isothermal
distribution with a core radius of \( 10'' \) (Eq.~\ref{eq:rhoiso}). These
two stellar distributions were chosen to represent the two cases of a sharp
density cusp and a flat core. We set the detection threshold at \( K_{0}=13\kmag  \),
corresponding to the OSIRIS data threshold, and find by trial and error that
the parameter values \( D=0.15 \), \( R=30\as  \) and \( A_{1}=2.5^{\mathrm{m}} \),
which correspond to a maximal extinction of \( 10.7^{\mathrm{m}} \), produce
a large enough spread in \( A_{K} \) at \( \sim 10\as  \). Dust models with
lower values of \( R \) and \( D \) fail to reproduce the highly reddened
stars because they do not have a high enough dust column, while in models with
higher values of \( R \) and \( D \) the extinction is so effective that the
central \( 10\as  \) is almost completely obscured. Figure \ref{fig:pak} show
that the models differ from the observations in several significant aspects.
Both predict a strong concentration of stars near and above \( A_{1} \), which
increases with \( p \) as the extinction decreases. Such an increase is not
observed in the data. Both also predict a nearly constant distribution of high-\( A_{K} \)
stars from \( p\sim 5\as  \) outwards (due to the fact that both the dust and
stellar columns fall together), unlike the observed distribution, which is confined
to the inner \( 10\as  \). The flat core model predicts very few high-\( A_{K} \)
stars in the inner \( \sim 5\as  \), since the flat stellar distribution does
not provide enough high luminosity stars that can be observed through the dust
cloud. This is to be contrasted with the dusty stellar cusp model, which does
predict a sparsely populated region in the \( p-A_{K} \) plane at \( p\la 10\as  \)
and \( A_{K}\la 7^{\mathrm{m}} \), approximately as is observed. However, the
stellar cusp model also predicts that a strong concentration of highly reddened
stars at \( p\la 2\as  \), contrary to the observations. We conclude that distributions
that are dusty enough to account for the highly reddened stars inside \( 10\as  \)
are not favored by the observed distribution of stars in the \( p-A_{K} \)
plane.

However, the \( p-A_{K} \) distribution in itself cannot rule out a combination
of a moderately dusty cloud together with a few intrinsically reddened stars,
and furthermore, our conclusions depend on the assumed stellar DF and the KLF.
The few spectra that were obtained for stars deep in the potential well can
provide additional information that is independent of these assumptions. Such
stars can be identified by their high velocities. Genzel et al. (\cite{Genzel96})
obtained spectra of 4 stars in the \( \SgrA  \) cluster (S1 at \( p=0.11'' \)
with projected velocity \( v_{\perp }=1556 \) km~s\( ^{-1} \), S2 at \( p=0.17'' \)
with \( v_{\perp }=691 \) km~s\( ^{-1} \), S8 at \( p=0.32'' \) with \( v_{\perp }=1086 \)
km~s\( ^{-1} \), and S11 at \( p=0.54'' \) with \( v_{\perp }=960 \) km~s\( ^{-1} \)).
Assuming that the stars are bound to the \( M_{\bullet }=2.6\times 10^{6}\, \Mo  \)
BH (Genzel et al. \cite{Genzel97}; Ghez et al. \cite{Ghez98}), these velocities
translate into maximal 3-dimensional distances from the center of \( r_{\mathrm{max}}=2G\Mbh /v_{\perp }^{2}=0.24'' \),
\( 1.2'' \), \( 0.48'' \), and \( 0.62'' \), respectively. The four spectra
are featureless and appear to be consistent with those of moderately reddened
(\( A_{K}\sim 3^{\mathrm{m}} \)) early-type stars in the central cluster (Genzel
et al. \cite{Genzel96}). This suggests that there is no significant concentration
of dust around the BH. 

We conclude that spherical dust distributions that are dusty enough to account
for the luminous, highly reddened stars in the inner \( 10\as  \) require a
stellar cusp to provide enough of such stars and are therefore inconsistent
with the absence of a strong concentration of highly reddened stars in the inner
few arcseconds. In addition, the absence of a color gradient in the inner \( 1\as  \)
and observations of moderately reddened stars deep in the potential well further
argue against a central dust cloud. On the other hand, the observed \( p-A_{K} \)
distribution is consistent with extinction by a uniform foreground dust screen,
provided that some intrinsically reddened stars exist in the inner \( 10\as  \).
Two such stars have in fact been detected there. A foreground dust screen will
not introduce a large systematic bias in the determination of the effective
power-law index of the stellar distribution.

\section{Stellar collisions and the depletion of luminous stars}

\label{sec:coll}

\begin{figure}[tbp]
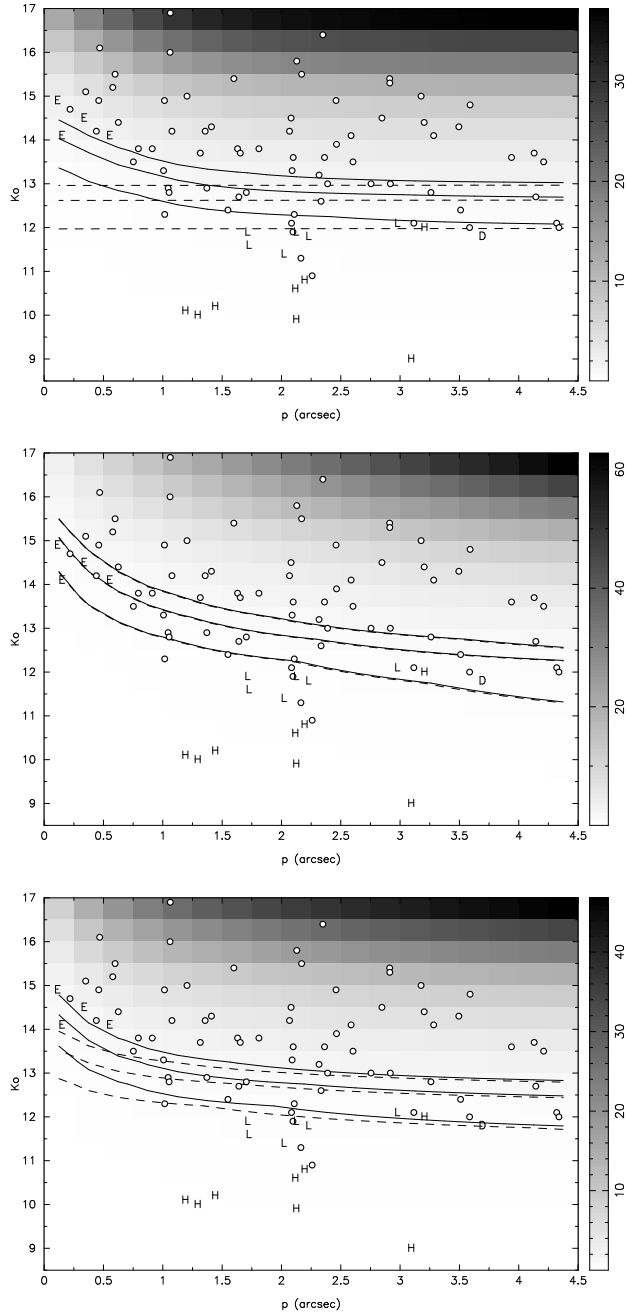

{\centering \begin{tabular}{c}
\resizebox*{!}{0.26\textheight}{\rotatebox{270}{\includegraphics{fig9a.ps}}} \\
\resizebox*{!}{0.26\textheight}{\rotatebox{270}{\includegraphics{fig9b.ps}}} \\
\resizebox*{!}{0.26\textheight}{\rotatebox{270}{\includegraphics{fig9c.ps}}} \\
\end{tabular}\par}

\caption{\label{fig:coll}The mean number of stars with \protect\( K<K_{0}\protect \)
per \protect\( 0.25\as \protect \)-wide bin, as expected with collisional destruction
of giants (gray-scale and full line contours) and without collisional destruction
(dashed line contours). The contour lines mark, from top to bottom, the model
predictions for the minimal magnitude below which less than 1.5, 1 and 0.5 stars
per \protect\( 0.25\as \protect \) bin are expected to be observed, on average.
Top panel: a \protect\( r^{-2}\protect \) stellar distribution. Middle panel:
a flat core in the inner \protect\( 10\as \protect \) and a \protect\( r^{-2}\protect \)
distribution outside. Bottom panel: a \protect\( r^{-3/2}\protect \) distribution
in the inner \protect\( 10\as \protect \) and a \protect\( r^{-2}\protect \)
distribution outside. Overlayed on the gray-scale are the stars observed by
Ghez et al. (\cite{Ghez98}), marked by different symbols according to their
spectral type (Eckart \& Genzel \cite{EG97}; Genzel et al. \cite{Genzel97}):
\HeI\noun{ }stars (H), late-type stars (L), early-type stars (E) , dust embedded
stars (D), and type unknown (circle).}
\end{figure}

The problem of the incompleteness of the counts could be circumvented, in principle,
by restricting the analysis to the brightest late-type stars, which should directly
reflect the projected stellar density. However, stars brighter than \( K=13.5\kmag  \)
are not observed in the inner \( 1\as  \) of the GC in the Keck data set, although
the counts are complete at least down to \( K=14\kmag  \). Figure~\ref{fig:coll}
shows the distribution of stellar magnitude against projected radius. Apart
from the counter-rotating \HeI \emph{\noun{}}stars, which clearly stand out
in the \( p-K \) plane, there is a gradual trend towards a fainter maximal
luminosity in the inner \( 2\as  \). The possibility that this is due to a
gradient in the extinction is not supported by the analysis in \S\ref{sec:Ak}.

Another interesting possibility is that the luminous stars have been destroyed
by stellar collisions. Stellar collisions were proposed as the mechanism for
destroying the outer envelopes of late-type giants\footnote{
The \( K \) luminosity of a star of radius \( R \) increases with the effective
temperature roughly as \( T_{\mathrm{eff}} \), so that the \( K \) luminosity
scales as \( R^{2}T_{\mathrm{eff}} \), while the bolometric luminosity scales
as \( R^{2}T^{4}_{\mathrm{eff}} \). A giant with \( R=100\, \Ro  \) and \( T_{\mathrm{eff}}=3\, 10^{3} \)
K that is reduced by a collision to \( \sim 1\, \Ro  \) will maintain its bolometric
luminosity by raising its effective temperature to \( T_{\mathrm{eff}}=3\, 10^{4} \)
K, but its \( K \) luminosity will fall by a factor of \( 10^{3} \) (\( 7.5\kmag  \)). 
} (Lacy, Townes \& Hollenbach \cite{LTH82}; Phinney \cite{Phinney89}) as an
explanation for the depletion of the CO absorption in the integrated light (Sellgren
\cite{Sellgren90}) or for the depletion of the bright late-giants in the counts
in the inner \( 5\as  \) of the GC (Genzel et al. \cite{Genzel96}). However,
the dynamical timescale in the envelope of a giant star is of the order of a
year and the thermal timescale is of the order of several decades, so that a
collision in which a star ``punches a hole'' in the giant's envelope will
probably not have a lasting effect. Studies of collisions between dwarf stars
and giants indicate that a significant fraction of the collisions can result
in a permanent destruction of the envelope if the ratio between the impact parameter
and the giant star radius, \( x_{e} \), is small enough, \( x_{e}\la 0.5 \).
The envelope is destroyed either by the ejection of the giant's core from the
extended envelope, which then rapidly disperses, or by the formation of a common
envelope binary, which stirs the envelope and leads to its evaporation on a
time scale much shorter than the giant phase (Livne \& Tuchman \cite{LT88};
Davies \& Benz \cite{DB91}; Rasio \& Shapiro \cite{RS90}, \cite{RS91}). These
studies concentrated on low mass \( \sim 1\Mo  \) giants and on low velocities
that are typical of globular clusters, so it is not quite clear how the results
scale to higher giant masses and velocities which are typical of the GC. More
recently, Davies et al. (\cite{Davies98}) proposed that 3-body collisions between
binaries and giants may provide a more efficient way of destroying the envelope.
They specifically tailored their dynamical simulations to the velocity field
and late-type giant masses and envelope sizes that are typical of the GC, and
found that envelope-destroying binary--giant collisions require \( x_{e} \)
of order unity. They also concluded that the role of such collisions is negligible
in the flat core of the flattened isothermal density model, even if all the
dwarf stars are in binaries. These works did not discuss collisional destruction
of early-type giant envelopes. However, these are likely to be less effective
since it appears that a necessary requirement is that the target star should
be composed of two distinct components, a compact core and an extended envelope
with a shallow density profile. 

In the following analysis we will assume that 2-body collisions result in the
permanent removal of the envelope of a late-type giant if \( x_{e}\le 0.25 \).
We estimate the efficiency of 3-body collisions by assuming that half of the
stars are in binaries. If envelope-destroying binary-giant collisions typically
require \( x_{e}=1 \), then the effective impact parameter \emph{per dwarf
star} is \emph{\( x_{e}=0.5 \)} (1 binary per 4 stars, \emph{}with the collisional
cross-section scaling as \( x^{2}_{e} \)) \emph{.} The effective impact parameter
per dwarf star must lie somewhere between the 2-body and 3-body values. Therefore,
by modeling both types of collisions as 2-body collisions, we are making a conservative
estimate of their effectiveness. 

Consider first a star on a circular orbit of radius \( r \) around the BH that
has total lifetime \( T \) and becomes a giant with a compact core / extended
envelope structure after a time \( t_{g} \). The Poisson probability for a
giant to survive envelope-destroying collisions with dwarf stars up to time
\( t>t_{g} \) is

\begin{equation}
\label{eq:survival}
S(t)=\exp \left( -\int ^{t}_{t_{g}}q(\tau )d\tau \right) \, ,
\end{equation}
where \( q \) is the local collision rate. The collision rate is the sum of
a geometrical cross-section term and a gravitational focusing term (see e.g.
Binney \& Tremaine \cite{BT87}),
\begin{equation}
\label{eq:rate}
q=4\sqrt{\pi }n_{p}\sigma (x_{e}R_{t}+R_{p})^{2}\left( 1+\frac{G(M_{t}+M_{p})}{2\sigma ^{2}(x_{e}R_{t}+R_{p})}\right) \, ,
\end{equation}
 where \( n_{p} \) is the number density of the dwarf projectiles, \( \sigma  \)
is the 1-dimensional velocity dispersion, and \( M_{t} \), \( M_{p} \), \( R_{t} \)
and \( R_{p} \) are the masses and radii of the giant targets and dwarf star
projectiles, respectively. Both terms have non-negligible contributions over
the wide range of stellar masses, radii and velocities in the GC (Eq.~\ref{eq:qg2qf}).
Equation~\ref{eq:rate} assumes a Maxwellian DF with an isotropic velocity dispersion.
These conditions appear to hold in the inner GC. The observed radial velocity
distribution is well fitted by a Gaussian distribution (Genzel et al. \cite{Genzel96})
and is thus consistent with theoretical predictions of the velocity distribution
very near a BH (Quinlan, Hernquist \& Sigurdsson \cite{Quinlan95}). The measured
velocities in the plane of the sky and along the line-of-sight are consistent
with an isotropic velocity dispersion (Ghez et al \cite{Ghez98}).

The collision rate \( q \) depends on the position \( r \) through \( n_{p} \)
and \( \sigma  \). The observed isotropic velocity dispersion in the inner
GC implies that stars observed at \( r \) have a wide range of orbital parameters
and spend a large fraction of their orbital period at distances much larger
or much smaller than \( r \). It is therefore necessary to take into account
the fact that the collision rates are not constant along the stellar orbits.
As we show in appendix~\ref{sec:orbit}, when the density and velocity dispersion
are power-laws in \( r \), the orbital averaging modifies Eq.~\ref{eq:rate}
only by introducing two constant correction factors for each of the geometrical
cross-section term and the gravitational focusing term. These factors depend
on the power-law indices and on the ratio \( \xi =v_{c}^{2}/\sigma ^{2} \),
where \( v^{2}_{c}=GM_{\bullet }/r \) is the circular velocity near the BH. 

In our analysis we assume that stars that are not bound to the BH are unaffected
by collisions, since the transit time through the inner \( 0.1 \)~pc is only
\( \sim 1000 \)~yr, which is much shorter than the typical mean time between
collisions (\( >10^{6} \) yr). Therefore, the orbit-averaged effective collision
rate depends very strongly on the value of \( \xi  \). The larger this ratio
is, the smaller the fraction of unbound stars, the more tightly bound the stars
whose orbits pass through \( r \) and the longer they spend in the high density
regions of the inner GC. The ratio \( \xi  \) is related to the stellar density
by the Jeans equation, which for a steady-state, non-rotating system with an
isotropic potential and isotropic velocity dispersion is
\begin{equation}
\xi \equiv \frac{v_{c}^{2}}{\sigma ^{2}}=-\frac{d\ln n}{d\ln r}-\frac{d\ln \sigma ^{2}}{d\ln r}\, .
\end{equation}
 Near the BH \( v^{2}_{c}\sim G\Mbh /r \), so the solution of the Jeans equation
for \( n\sim r^{-\alpha } \) is \( \sigma ^{2}\sim 1/r \) and therefore \( \xi \sim 1+\alpha  \). 

Genzel et al. (\cite{Genzel96}) assumed a flattened power-law density model
(\S\ref{sec:3D+SHARP}), which is flat (\( \alpha \sim 0 \)) near the BH, and
used the Jeans equation to estimate \( \xi  \) from the observed projected
velocity dispersion \( \sigma _{p} \). They assumed a deprojected velocity
dispersion of the form \( \sigma ^{2}=\sigma ^{2}_{\infty }+\sigma _{c}^{2}(r/r_{c})^{-\gamma } \)
with \( \sigma _{\infty } \) a fixed constant, and obtained a best fit value
of \( \gamma =1.2 \) and consequently a \( \xi =v_{c}^{2}/\sigma ^{2} \) ratio
that falls from \( \sim 2 \) at \( r\sim r_{c} \) down to \( 1.2 \) at \( r=0 \).
However, their velocity dispersion model is inconsistent with a density cusp.
For example, the Jeans equation associates a \( n\propto r^{-2} \) cusp with
a smaller deprojected velocity dispersion that varies as \( \sigma ^{2}=G\Mbh /3r \).
This can be understood qualitatively by noting that in a flat core the stars
have a wider distribution along the line-of-sight and are on average farther
away from the BH than they are in a cusp. Therefore, in a flat core a large
value of \( \sigma _{p} \) implies that the stars move at a large fraction
of the circular velocity, whereas in a cusp this merely reflects the fact that
a larger fraction of the stars are deep in the potential well. A reduction of
the velocity dispersion significantly changes the orbital structure of the population.
The fraction of stars on orbits more bound than circular (\( v^{2}<v_{c}^{2} \)),
less bound than circular (\( v_{c}^{2}\leq v^{2}<2v_{c}^{2} \)) and unbound
(\( 2v_{c}^{2}\leq v^{2} \)) are respectively \( 0.20:0.23:0.57 \) for \( \xi =1 \),
\( 0.43:0.31:0.26 \) for \( \xi =2 \), and \( 0.61:0.28:0.11 \) for \( \xi =3 \).
That is, the steeper the cusp, the more the stars are bound to the black hole.
This implies that the stellar population in a flat core is significantly more
mixed than it would be in a cusp, and that any local spectral variations in
the stellar population will tend to be diluted. 

The probability that a star observed at a random time \( t \) during its lifetime
will be brighter than some magnitude \( K_{0} \) is given by

\begin{equation}
P_{0}=\frac{1}{T}\left( \int ^{t_{g}}_{0}m_{0}(t)dt+\int ^{T}_{t_{g}}m_{0}(t)S(t)dt\right) \, ,
\end{equation}
where \( m_{0}(t) \) is a function that equals 1 when the star is brighter
than \( K_{0} \), and equals 0 otherwise. The probability \( P_{0} \) depends
on the star's evolutionary track through the fraction of time the star spends
in the giant phase and through the evolution of the stellar radius and the \( K \)-luminosity.
The probability \( P_{0} \) has to be averaged over the stellar population
to obtain the total probability of observing any star brighter than \( K_{0} \).
As was discussed in \S\ref{sec:KLF}, the stellar population in the GC that
contributes to the KLF is well modeled by a continuous star-forming history
with an approximately Salpeter initial mass function with a minimal mass of
\( M_{1}=0.8\Mo  \) and a maximal mass of \( M_{2}=120\Mo  \), so that the
present-day mass function (PMF) can be approximated as 
\begin{equation}
\frac{df}{dM_{0}}(t_{0})\propto \Psi \left\{ \begin{array}{lr}
M^{-2.35}_{0}T & T<t_{0}\\
M^{-2.35}_{0}t_{0} & T\geq t_{0}
\end{array}\right. \, ,
\end{equation}
where \( \Psi  \) is the star-formation rate, which we take to be constant,
\( M_{0} \) is the zero-age main sequence stellar mass, and \( t_{0} \) is
the age of the system, which we take to be 10~Gyr.

We calculate the PMF-averaged \( P_{0} \) using the Geneva stellar evolution
tracks (Schaerer et al. \cite{Schaerer93}) for twice-solar metallicity stars
and assume blackbody stellar spectra for tracking the \( K \) luminosity with
time. We model the stellar DF by a two power-law function (Eq.~\ref{eq:rhopl})
with \( r_{b}=10\as  \) and \( \rho _{b}=(3-\alpha )\times 10^{6}\Mo \mathrm{pc}^{-3} \),
where the pre-factor normalizes all models to have an equal mass of \( \sim 8\times 10^{5}\, \Mo  \)
within \( r_{b} \). For the dwarf projectiles (burning stars and stellar remnants),
we assume \( n_{p}(r_{b})=\rho _{b}/\left\langle m\right\rangle  \) with \( \left\langle m\right\rangle =0.6\Mo  \),
while for the target stars (burning stars only), we assume \( n_{t}(r_{b})=\rho _{b}/4 \)
(see Alexander \& Sternberg \cite{AS99}). The number of stars per \( p \)-annulus
that are brighter than \( K_{0} \) is then given by 

\begin{equation}
N(p;K_{0})=2\pi p\Delta p\int _{-\infty }^{+\infty }\int _{M_{1}}^{M_{2}}n_{t}(p,z)\left[ P_{0}(1-f_{ub})+f_{ub}\right] \frac{df}{dM_{0}}dM_{0}dz\, ,
\end{equation}
 where \( f_{ub} \) is the fraction of unbound stars in the population and
where \( P_{0} \) is calculated with the effective orbit-averaged collision
rate as described in appendix \ref{sec:orbit}. 

We begin by considering a \( r^{-2} \) cusp (\( \xi =3 \) and \( f_{ub}=0.11 \)).
Figure \ref{fig:coll} overlays the distribution of the observed stars on a
binned gray-scale plot of \( N(p;K_{0}) \). The middle contour line shows the
model prediction for the values of the minimal magnitude above which less than
one star per \( p \)-bin is expected to be observed, with and without collisions.
The \( n\propto r^{-2} \) cusp model with collisions reproduces the observed
trend to a remarkable degree, given that no attempt was made to fit the data.
Apart for the \noun{}\HeI stars, which clearly constitute a separate sub-population,
and the other young stars (early-type and dust-embedded stars), there are in
total 24 stars below the `1 star per \( p \)-bin' contour. This is statistically
consistent with the expected value of 18 for the 18 \( p \)-bins between \( 0\as  \)
and \( 4.5\as  \), especially given the possibility that a few of the 24 stars
could be unidentified early type stars. The minimal magnitude contour for the
no-collisions case clearly fails to reproduce the observations, as it remains
flat down to the smallest projected radii. Our model for the collisional depletion
of giants predicts that the collisions become effective only at \( p\la 2\as  \),
which corresponds to \( n_{p}>4\times 10^{7}\, \mathrm{pc}^{-3} \) and \( \sigma >300\, \mathrm{km}\, \mathrm{s}^{-1} \). 

Up to this point we considered only the collisionless orbits of the stars. However,
two-body encounters will also affect the orbits. The typical mean time between
envelope-destroying collisions in the inner \( 2\as  \) is \( \sim 10^{7} \)~yr
per giant, whereas the crossing time of the collisionally-depleted region is
only \( \sim 500 \)~yr. This raises the question whether the central region
can avoid being replenished by collisionally-deflected giants from the dense
stellar environment just outside it. A detailed calculation of the replenishment
rate is beyond the scope of this work, but we note that there are several reasons
that suggest that this process is likely to be inefficient. Slow diffusion of
giants into the inner region may make them as vulnerable to collisional destruction
as the giants that were originally bound to that region. Giants that will be
deflected by a single close elastic collision into a low angular momentum orbit
will spend only a small fraction of their lifetime in the central \( 2\as  \).
Moreover, the velocity dispersion in the cusp is high, so that large deflection
angles require very close encounters. This defines a collisional radius, \( r_{\mathrm{coll}} \)
(Frank \& Rees \cite{Frank76}), inside which large deflections require physical
(inelastic) collisions. For a \( R_{t}=100\, \Ro  \), \( M_{t}=3\, \Mo  \)
giant, the collisional radius is

\begin{equation}
r_{\mathrm{coll}}=\frac{\Mbh }{M_{t}}R_{t}\sim 2\, \mathrm{pc}\sim 50\as \, .
\end{equation}
Therefore, a close encounter within \( x_{e}r_{\mathrm{coll}} \), which is
close enough to deflect the giant into a low angular momentum orbit, will also
destroy its envelope. Replenishment by giants from beyond \( x_{e}r_{\mathrm{coll}} \)
will be less effective because of the lower stellar density there and because
of the small angular size of the central \( 2\as  \), as seen from \( r>x_{e}r_{\mathrm{coll}} \). 

Next, we consider the case of a flat core (\( \alpha =0 \), \( \beta =2 \),
\( r_{b}=10\as  \) with \( \xi =1 \) and \( f_{ub}=0.57 \)). Surprisingly,
the flat core model shows that the apparent depletion of the high luminosity
stars can also be a \emph{projection} effect. Figure \ref{fig:coll} shows that
collisions play no role in such a low density models, in agreement with the
conclusion of Davies et al. (\cite{Davies98}) but contrary to the suggestion
of Genzel et al. (\cite{Genzel96}). However, since the number of stars per
\( p \)-annulus falls as \( p \) in a flat core, the probability of observing
one of the rare luminous star also falls with \( p \). This is to be contrasted
with the \( r^{-2} \) cusp, where the number of stars per annulus is constant. 

The origin of the depletion in a \( r^{-2} \) cusp is collisional, while in
a flat core it is due to a projection effect\footnote{
A \( r^{-\alpha } \) cusp with \( \alpha >2 \) will have collisional depletion
offset by apparent \emph{enhancement} due to projection.
}. Stellar cusps shallower than \( r^{-2} \) will display both types of depletion.
Close inspection of the predicted depletion in the \( r^{-2} \) cusp (Fig.~\ref{fig:coll})
shows that it somewhat underestimates the depletion at the smallest projected
distances, and that a better fit can be obtained by a shallower cusp, for example
\( r^{-3/2} \). 

We conclude that the depletion of the bright stars in the inner \( 2\as  \)
can be explained by a combination of collisional destruction of giant envelopes,
which requires the high densities of a stellar cusp, and by projection effects.
The latter is due to the fact that density distributions that are shallower
than a \( r^{-2} \) cusp have progressively less stars enclosed in smaller
projected radii, and so there is a smaller chance of observing the rarer, more
luminous stars there. Because of the the degeneracy between the collisional
depletion and the projection effects, the distribution of the brightest stars
alone cannot decide the case between a flat core and a cusp.

\section{Discussion}

\label{sec:discussion}

\subsection{The star counts}

The analysis of the star counts is not straightforward and must overcome several
difficulties. First, the observed star-counts have yet to be published in a
way suitable for detailed analysis by taking into account completeness corrections
and field edge corrections. Second, the GC contains a population of luminous
young stars, which are not dynamically relaxed, and therefore introduce noise
in the counts. Third, a gradient in the extinction field may bias the counts.
Although none of these concerns can be completely eliminated, we have attempted
to minimize their possible effects in our analysis.

An important aspect of our effort to control the effects of incompleteness was
the use of three different data sets, with different properties and biases.
We analyzed them in a uniform way by methods that avoid binning, as is appropriate
when dealing with sparse data. It is reassuring that all three data sets indicate
that a cusp matches the counts better than a flat core. The case for a cusp
is further strengthened by the fact that both incompleteness due to source confusion
and a central concentration of dust will tend to suppress the observed cusp.
Another way of circumventing the completeness problem is to consider the information
in the distribution of the brightest stars. We discuss this further below.

We attempted to limit the contamination by the young unrelaxed population by
excluding all the identified young stars from the data sets. Although we do
not know how many of the fainter stars are associated with this unrelaxed population,
the star-burst model of Krabbe et al. (\cite{Krabbe95}) predicts that their
fraction should become negligible as the counts go deeper. It is possible that
this trend is already apparent in the differences between the OSIRIS data set
and the two other deeper data sets (Figure~\ref{fig:cum}). The ML power-law
index does not change significantly when the young stars are included in the
sample (Figure~\ref{fig:ML}). 

To estimate the possible effects of extinction gradients, we constructed a few
simple 3-dimensional dust distribution models and tested them against the observed
distribution of stellar extinction in the projected distance--reddening plane.
We find that on the \( \sim 10\as  \) scale, the observations are consistent
with an external dust screen, which is unlikely to significantly bias the derived
power-law index of the stellar distribution. Our constraints on variations on
\( \sim 1\as  \) scales are weaker, but we note that the detection of moderately
reddened stars very deep in the potential well is consistent with little reddening
in the innermost GC, and that no significant color gradient was detected in
the inner \( 1\as  \). 

These precautions and checks are no substitute for better data. Clearly, deep,
completeness-corrected data sets are desirable. A comparison of the likelihood
curves of the Keck and 3D+SHARP data set shows that even a two-fold increase
in the sample size from \( \sim 50 \) to \( \sim 100 \) can lead to a significant
improvement in the ability to discriminate between stellar distribution models.
Our KLF model suggests a two-fold increase in the number of stars observed for
every \( 1\kmag  \) improvement in the photometric sensitivity.

\subsection{The depletion of bright stars }

In addition to the analysis of the total star counts, we examined the distribution
of the counts as function of magnitude and noted that the highest resolution
data set shows a gradual depletion of the brightest stars in the inner \( 2\as  \)
(a reflection of the fact that there are no bright stars in the \( \SgrA  \)
cluster). Reproducing this effect by reddening would require a \( \sim 2.5\kmag /2\as  \)
extinction gradient, which is not indicated by our analysis (\S\ref{sec:Ak}).
We showed that envelope-destroying collisions between dwarfs and giants can
reproduce the observed trend to a remarkable degree, once the young unrelaxed
stars are excluded from the data set. Our detailed model for the effects of
collisions on the luminosity function included six components: realistically
lowered estimates for the cross-section of envelope-destroying collisions, based
on results from hydrodynamical simulations of dwarf-giant collisions; an orbit-averaged
estimate of the collision rate; a velocity dispersion model that is self-consistent
with the assumed density distribution; a stellar mass function that reproduces
the mean observed \( K \) luminosity function and mass-to-light ratio; stellar
tracks that model the evolution of the size and luminosity of the stars as function
of their initial mass; and a careful statistical treatment of the survival probability. 

Our dynamical analysis shows that a cusp is necessary for collisional depletion
to play a role because the process becomes effective only at a density of order
\( >5\times 10^{7}\Mo \mathrm{pc}^{-3} \) with a velocity dispersion of order
\( >300\, \mathrm{km}\, \mathrm{s}^{-1} \) , and because a cusp is much more
tightly bound to the BH than a flat core. The unbound stars spend too little
time in the high density central regions to be affected by the collisions, and
so set an upper limit on the fraction of giants in the population that can be
effectively destroyed. 

Stellar collisions are only one way to explain the absence of bright stars without
having to invoke intrinsic variations in the stellar population. We also find
that a flat core can qualitatively mimic such a depletion by a projection effect,
even when collisions play no role. This explanation for the absence of bright
stars in a flat core may actually be more likely than to assume intrinsic variations
in the stellar population. Any localized, distinct stellar population in a flat
core will be diluted by the large fraction of loosely bound stars that sample
the population from over the entire GC.

\subsection{Implications of a stellar cusp in the GC}

We now discuss briefly some of the implications of the existence of a steep
stellar cusp in the GC. The tidal disruption rate of stars by the BH is enhanced
in a steep cusp. Such events may be observed as short and luminous flares (Rees
\cite{Rees88}). Bahcall \& Wolf (\cite{BW77}) estimated the diffusion rate
into the `loss-cone' (low angular-momentum orbits) as function of the BH mass
and of the line-of sight velocity dispersion and the stellar density outside
the cusp. Adopting a broken power-law density distribution (Eq.~\ref{eq:rhopl})
with \( r_{b}=10\as  \), \( \beta =2 \) and \( n_{b}=10^{6}\, \mathrm{pc}^{-3} \)
and taking \( \sigma _{p}\sim 85\, \mathrm{km}\, \mathrm{s}^{-1} \) at \( \sim 2.5 \)~pc
from Kent (\cite{Kent92}) or \( \sigma _{p}\sim 60\, \mathrm{km}\, \mathrm{s}^{-1} \)
at \( \sim 4 \)~pc from Genzel et al. (\cite{Genzel96}), we estimate a loss-cone
diffusion rate of \( \mathrm{few}\times 10^{-5}\, \mathrm{yr}^{-1} \). This
is consistent with the rates derived by Magorrian \& Tremaine (\cite{MT99})
for faint galaxies (\( L<10^{10}\, L_{\odot } \)) with low-mass BHs and steep
density cusps, but is much larger than the \( 10^{-9}-10^{-8}\, \mathrm{yr}^{-1} \)
rates that they derived for \( L^{\star } \) galaxies. Magorrian \& Tremaine
suggest that tidal flares are most likely to be detected in faint galaxies.
Our results raise the possibility that at least some luminous galaxies may have
tidal flare rates as high as those of faint galaxies.

Inelastic dwarf--dwarf collisions are expected to modify the stellar distribution
in the inner, highest density regions of the cusp by gradually destroying the
tightly bound stars. This leads to the formation of a \( n\propto r^{-1/2} \)
cusp of unbound stars. The relevant time-scale for this process is the time
it takes to destroy a star by successive collisions. Murphy, Cohn and Durisen
(\cite{MCD91}) show that the collisional destruction timescale is \( \gtrsim 10t_{\mathrm{coll}} \),
where \( t_{\mathrm{coll}} \) is the collision timescale per star. The size
of the unbound cusp that they derive in two of their numerical models that roughly
bracket the conditions in the GC (models 3B and 4B), is \( 0.002<r_{d}<0.04 \)
pc at 15 Gyr. This qualitatively agrees with our independent estimate of \( t_{\mathrm{coll}}\lesssim 5\times 10^{8} \)
yr at \( r<0.02 \) for dwarf--dwarf collisions in a \( \alpha =1.5 \) cusp.
We therefore expect the cusp to flatten somewhere within \( p\lesssim 0.5\as  \).
We estimate that dwarf--dwarf inelastic collisions occur in that volume at a
rate of \( 10^{-5}-10^{-4}\, \mathrm{yr}^{-1} \).

A steep stellar cusp will also increase the rate of microlensing of stars by
the BH (Alexander \& Sternberg \cite{AS99}), and especially that of short duration
microlensing events of stars very near the BH. Such microlensing light-curves
can be used, in principle, to probe the mass distribution of the central dark
mass and test whether it is a point mass. These issues will be discussed in
more detail elsewhere.

Finally, we note that the integrated stellar mass close to the black hole is
very small in spite of the very high density of the cusp. For example, the stellar
mass within \( 2\as  \) in a \( r^{-2} \) cusp is \( \sim 8\times 10^{4}\, \Mo  \)
as compared to only \( \sim 8\times 10^{3}\, \Mo  \) in the flattened isothermal
distribution (when both are normalized to have the same mass within \( r_{c}=10\as  \)).
This difference amounts to only a \( \sim 3\% \) over-estimate in the BH mass
due to the fact that some of the stellar mass is erroneously attributed to the
BH.

\section{Summary}

\label{sec:summary}

Recent deep high-resolution observations of the inner GC and large scale multi-color
observations have supplied a wealth of information on the spatial distribution
of stars near the super-massive BH in the GC. This offers the opportunity to
map the stellar distribution around the BH directly from the star counts, rather
than from the integrated light. Dynamical scenarios predict that a stellar cusp
should form around the BH, and that the shape of the cusp could provide clues
on the formation history of the BH and the stellar system around it. It is therefore
puzzling that recent empirical GC mass models indicate a flat core rather than
a cusp, while at the same time, deep observations reveal an over-density of
faint stars and an absence of bright stars in the inner \( 1\as  \) (the \( \SgrA  \)
cluster). Taken at face value, this picture raises several questions: What does
the absence of a cusp imply about the dynamical state of the GC? Are the faint
stars a distinct population? If so, how have they formed in the inner \( 1\as  \)
or how have they migrated there? Our analysis of three independent sets of star
counts points to a simpler interpretation of these findings. Our main results
are as follows.

\begin{enumerate}
\item The central over-density is likely the tip of a stellar cusp that rises smoothly
in the inner \( \sim 10\as  \). A stellar cusp is consistent with the star-counts
and is preferred at the \( \sim 2\sigma  \) level over a flat core distribution
even by the least restrictive data set. A flat core is completely ruled out
by the largest data set. The maximum likelihood range for the power-law index
of the cusp is \( \alpha \sim 1.5 \) to \( 1.75 \). An alternative way of
stating this result is that our analysis points to a small core radius, \( r_{c}<2.5\as  \).
\item The gradual depletion of the luminous giants in the inner \( 2\as  \) can be
explained by two effects: envelope-destroying dwarf-giant collisions in a steep,
dense and tightly bound cusp, or a projection effect in a flatter distribution,
where the number of stars per projected annulus falls with decreasing projected
radius. The observed depletion in the innermost regions is better described
by a \( \alpha <2 \) cusp, where both effects contribute. Either way, it is
unnecessary to assume that the stellar population in the innermost GC is intrinsically
different (fainter) from the general population. 
\item The \emph{joint constraints} of the star counts and the depletion of the brightest
stars suggest that a power-law cusp with \( \alpha  \) in the range \( 3/2 \)
to \( 7/4 \) provides a good overall description of the data. Such a cusp is
consistent with the Bahcall-Wolf solution (Bahcall \& Wolf \cite{BW76}, \cite{BW77};
Murphy, Cohn \& Durisen \cite{MCD91}; see \S\ref{sec:theory}) for the distribution
of intermediate mass stars in a multi-mass stellar system that has undergone
two-body relaxation around a black hole (Eq.~\ref{eq:mslope}). Dynamical estimates
(Eq.~\ref{eq:trelax}) suggest that the inner GC is relaxed. If future deeper
star counts support these conclusions, this would be the first detection of
such a system.
\end{enumerate}
\acknowledgments

I am grateful to A. Eckart, R. Genzel and A. Ghez for supplying additional information
about their data and to J. Bahcall, D. Eisenstein, J. Goodman and S. Tremaine
for helpful comments.

\appendix

\section{The stellar ``ring'' at \protect\( 2''\protect \)}

\label{sec: ring}Is the striking over-concentration of stars in a thin ring
at \( p\sim 2'' \), which is seen in the cumulative distribution of the Keck
data set (Figs.~\ref{fig:ghez} and \ref{fig:cum}), inconsistent with a power-law
distribution (9 late-type stars in a ring of width \( 0.06\as  \) out of a
total of 54 late-type star or 11 stars of all spectral types out of a total
of 64 stars)? This is probably not an instrumental artifact, since it is seen
also in the full OSIRIS data set. The 11 stars do not seem to have any other
peculiar characteristics apart from their spatial alignment. Only two (16CC
and 29N) are identified as \HeI stars (Eckart \& Genzel \cite{EG97}), their
mean magnitude of \( \overline{K}=12.6^{\mathrm{m}}\pm 2.9^{\mathrm{m}} \)
is consistent with the total average of \( \overline{K}=13.4^{\mathrm{m}}\pm 2.8^{\mathrm{m}} \),
and their projected velocity vectors appear to be random. In order to address
this question, we performed Monte-Carlo simulations where we randomly drew 64
stars from a \( n\propto r^{-\alpha } \) distribution and recorded the number
of times a group of \( k \) stars or more was concentrated within a ring of
\( \Delta p=0.06\as  \) or less, for various values of \( k \) and \( \alpha  \)
(a ring of \( k \) stars is counted twice if it is a subset of a ring of \( k+1 \)
stars). The distribution of the ring radii was also recorded. The results, listed
in table~\ref{tab: ring}, show that a ring of 11 stars in \( 0.06\as  \) is
quite rare in a \( \alpha =2 \) model, where there are on average only \( 2.3\, 10^{-5} \)
such rings per random realization, with radii \( 1.3\as \pm 0.7\as  \), but
less rare in the flatter \( \alpha =3/2 \) model, where the average number
is \( 2.3\times 10^{-4} \) with radii \( 2.1\as \pm 0.3\as  \). The probabilities
increase by more than an order of magnitude if only the 9 late-type stars are
considered members in the ring.

The Monte-Carlo averaged ring position reflects a trade-off between the ring
area, which increases with \( p \), and the density distribution, which decreases
with \( p \). Therefore, both the probability for observing a ring, and the
ring's radius, increase in flatter distributions. The fact that an 11-star ring
of radius \( 2\as  \) is observed, can be turned around and used as a Maximum
Likelihood argument in favor of a flatter inner distribution. Future deeper
number counts will show whether this feature is more than a statistical fluctuation.

\section{The effective orbit-averaged collision rate}

\label{sec:orbit}

We estimate the corrections to the local collision rate (Eq.~\ref{eq:rate})
due to the fact that the stars are not on circular orbits by making two simplifying
assumptions. First, we assume that the stars in the inner GC move on Keplerian
orbits around the BH, so that the orbits can be parameterized by their semi-major
axis \( a \) and their eccentricity \( e \). Second, we assume that unbound
stars are unaffected by collisions (see \S\ref{sec:coll}). When the orbital
period \( P \) of a star around the BH is shorter than the giant phase, the
orbit-averaged collision rate is 
\begin{equation}
\bar{q}(r,a,e)=\frac{2}{P}\int ^{a(1+e)}_{a(1-e)}q(s)\frac{ds}{\dot{s}}\, ,
\end{equation}
where \( s \) is the radial coordinate of the orbit and \( a(1\pm e) \) are
the apoastron and periastron radii, respectively. The radial velocity varies
with \( s \) as 
\begin{equation}
\dot{s}^{2}=2\left( \epsilon +\frac{GM}{s}\right) -\frac{h^{2}}{s^{2}}\, ,
\end{equation}
where \( \epsilon =-GM_{\bullet }/2a \) is the specific energy and \( h=\sqrt{GM_{\bullet }a(1-e^{2})} \)
is the specific angular momentum of the star. The orbital parameters of a star
at \( r \) with velocity \( v \) can be expressed in terms of \( \tilde{v}=v/\sigma  \)
and the angle between the radius and the velocity vectors, \( \eta  \), 
\begin{equation}
a=\frac{r}{2-\tilde{v}^{2}/\xi }\, ,
\end{equation}
\begin{equation}
e^{2}=1+\sin ^{2}\eta \left( \frac{\tilde{v}^{4}}{\xi ^{2}}-\frac{2\tilde{v}^{2}}{\xi }\right) \, ,
\end{equation}
where \( \xi =v^{2}_{c}/\sigma ^{2} \). For \( n\propto r^{-\alpha } \) and
\( \sigma ^{2}\propto r^{-\gamma } \), the geometrical cross-section term and
the gravitational focusing term in the local collision rate can be expressed
as \( q(r)=q_{g}+q_{f}=Ar^{-(\alpha +\gamma /2)}+Br^{-(\alpha -\gamma /2)} \),
where the constants \( A \) and \( B \) are defined by Eq.~\ref{eq:rate}.
In this case, the corresponding orbit-averaged terms are
\begin{equation}
\bar{q}_{g,f}(r)=\left[ (2-\tilde{v}^{2}/\xi )^{(\alpha \pm \gamma /2)}\frac{\sqrt{x_{1}x_{2}}}{\pi }\int ^{x_{2}}_{x_{1}}\frac{x^{\alpha \pm \gamma /2-2}dx}{\sqrt{(x-x_{1})(x_{2}-x)}}\right] q_{g,f}(r)\equiv C_{g,f}(\tilde{v},\eta )q_{g,f}(r)\, ,
\end{equation}
where \( x_{1,2}=1/(1\pm e) \) and where the \( (+) \) case corresponds to
\( \bar{q}_{g} \) and the \( (-) \) case to \( \bar{q}_{f} \). It then follows
that \( \left\langle \bar{q}\right\rangle _{g,f} \), the averages of \( \bar{q}_{g,f} \)
over the Maxwellian DF of the bound stars, are related to the local collision
rates \( q_{g,f} \) by proportionality factors that are functions of the constants
\( \alpha  \), \( \gamma  \) and \( \xi  \) but not of \( r \),

\begin{equation}
\label{eq:Q}
\left\langle C\right\rangle _{g,f}\equiv \left\langle \bar{q}\right\rangle _{g,f}/q_{g,f}=(2\pi )^{-1/2}\int ^{\sqrt{2\xi }}_{0}\int ^{\pi }_{0}C_{g,f}(\tilde{v},\eta )\tilde{v}^{2}\exp \left( -\tilde{v}^{2}/2\right) \sin \eta \, d\tilde{v}d\eta \, .
\end{equation}
In practice, the expectation value \( \left\langle C\right\rangle  \) diverges
for certain choices of \( \alpha  \) and \( \gamma  \) due to low angular-momentum
orbits that sample regions of diverging density and velocity dispersion. For
this reason, and because the survival probability (Eq.~\ref{eq:survival}) is
a highly non-linear function of the collision rate, a more relevant quantity
is the \emph{median} correction factor \( \widehat{C} \), defined as the ratio
between the \emph{}median \emph{}of \( \bar{q} \) over the population of bound
stars and the local rate \( q \). Table~\ref{tab:orb} lists \( \widehat{C}_{f} \)
and \( \widehat{C}_{g} \) for different density distributions that may be relevant
in the inner GC, as well as the fraction of unbound stars, \( f_{ub} \), that
corresponds to each value of \( \xi  \). The median correction factors were
estimated by \( 10^{6} \) Monte Carlo draws of random velocity vectors from
the DF. Using these factors, the effective collision rate for the bound stars
is estimated by
\begin{equation}
\label{eq:qeff}
q_{\mathrm{eff}}=A\widehat{C}_{g}r^{-(\alpha +\gamma /2)}+B\widehat{C}_{f}r^{-(\alpha -\gamma /2)}\, .
\end{equation}
 The relative contributions of the geometrical cross-section term and the gravitational
focusing term to the effective collision rate is (cf. Eq.~\ref{eq:rate})
\begin{equation}
\label{eq:qg2qf}
\left( \frac{q_{g}}{q_{f}}\right) _{\mathrm{eff}}=\frac{\widehat{C}_{g}}{\widehat{C}_{f}}\frac{2}{\xi }\left( \frac{\Mbh }{M_{t}+M_{p}}\right) \left( \frac{x_{e}R_{t}+R_{p}}{r}\right) \, .
\end{equation}
 This ratio can be both much larger than 1 or smaller than 1 over the wide range
of stellar masses, radii and distances in the GC, so that both terms should
be taken into account when calculating \( q_{\mathrm{eff}} \).

\clearpage %Figure captions


\clearpage 

 \clearpage %Tables

\begin{table}[tbp]

\caption{\label{tab: ks}The Kolmogorov-Smirnov acceptance probabilities for various
stellar density models. The broken power-law models have \protect\( r_{b}=10\as \protect \)
and the flattened isothermal model \protect\( r_{c}=10\as \protect \). The
highest probabilities for each data set are emphasized in italics.}
{\centering \begin{tabular}{ccccc}
\hline 
data set &
Keck &
Keck &
OSIRIS &
3D+SHARP\\
Subset&
All&
old stars &
old stars&
old stars w/o \( \SgrA  \)\\
No. of stars&
64&
54&
50&
107\\
\hline 
Flat. isothermal&
0.05&
0.12&
0.41&
0.20\\
\( r^{-2} \)&
0.02&
0.02&
0.10&
0.04\\
\( r^{-7/4} \)&
0.11&
0.12&
0.78&
\emph{0.63}\\
\( r^{-3/2} \)&
\emph{0.38}&
\emph{0.3}7&
0.33&
0.50\\
\( r^{-7/4}+r^{-2} \)&
0.09&
0.10&
0.51&
0.34\\
\( r^{-3/2}+r^{-2} \)&
0.27&
0.27&
\emph{0.91}&
0.56\\
\hline 
\end{tabular}\par}\end{table}

%\clearpage

\begin{table}[tbp]

\caption{\label{tab: ring}The frequency of ring-like structures in \protect\( 10^{7}\protect \)
Monte-Carlo realizations drawn from a \protect\( n\propto r^{-\alpha }\protect \)
density cusp. In each realization, 64 stars are randomly distributed within
\protect\( 2.5''\protect \), and the number of rings of width \protect\( \leq 0.06\as \protect \)
is recorded as function of the number of stars in the ring.}
{\centering \begin{tabular}{rrr}
\hline 
\multicolumn{1}{c}{Stars per ring}&
\multicolumn{1}{c}{Rings per realization}&
\multicolumn{1}{c}{Mean ring position}\\
\hline 
\multicolumn{3}{l}{\( \alpha =3/2 \)}\\
\( \geq 7 \)&
\( 1.8\, 10^{-1} \)&
\( 2.0''\pm 0.4'' \)\\
\( \geq 8 \)&
\( 4.0\, 10^{-2} \)&
\( 2.0''\pm 0.4'' \)\\
\( \geq 9 \)&
\( 8.1\, 10^{-3} \)&
\( 2.1''\pm 0.3'' \)\\
\( \geq 10 \)&
\( 1.5\, 10^{-3} \)&
\( 2.1''\pm 0.3'' \)\\
\( \geq 11 \)&
\( 2.3\, 10^{-4} \)&
\( 2.1''\pm 0.3'' \)\\
\multicolumn{3}{l}{\( \alpha =2 \) }\\
\( \geq 7 \)&
\( 5.4\, 10^{-2} \)&
\( 1.3\as \pm 0.7\as  \)\\
\( \geq 8 \)&
\( 9.2\, 10^{-3} \)&
\( 1.3\as \pm 0.7\as  \)\\
\( \geq 9 \)&
\( 1.4\, 10^{-3} \)&
\( 1.3\as \pm 0.7\as  \)\\
\( \geq 10 \)&
\( 1.9\, 10^{-4} \)&
\( 1.3\as \pm 0.7\as  \)\\
\( \geq 11 \)&
\( 2.3\, 10^{-5} \)&
\( 1.3\as \pm 0.7\as  \)\\
\hline 
\end{tabular}\par}\end{table}

%\clearpage

\begin{table}[tbp]

\caption{\label{tab:orb}The orbital correction factors to the local collisional rate
(Eq.~\ref{eq:qeff}) and the fraction of unbound stars in the population.}
{\centering \begin{tabular}{cccccc}
\hline 
\( \alpha  \)&
\( \gamma  \)&
\( \widehat{C}_{g} \)&
\( \widehat{C}_{f} \)&
\( \xi =\alpha +\gamma  \)&
\( f_{ub} \)\\
\hline 
5/2&
1&
6.2&
2.3&
7/2&
0.07\\
2&
1&
3.5&
1.7&
3&
0.11\\
7/4&
1&
2.8&
1.5&
11/4&
0.14\\
3/2&
1&
2.2&
1.4&
5/2&
0.17\\
1&
1&
1.6&
1.2&
2&
0.26\\
1/2&
1&
1.1&
1.1&
3/2&
0.39\\
0&
1&
0.9&
1.0&
1&
0.57\\
\hline 
\end{tabular}\par}\end{table}


\clearpage

\begin{thebibliography}{1999}
\bibitem[1999]{AS99}Alexander, T., \& Sternberg, A., 1999, ApJ, 520, in press
\bibitem[1983]{Allen83}Allen, D. A., Hyland, A. R., \& Jones, T. J., 1983, MNRAS, 204, 1145
\bibitem[1976]{BW76}Bahcall, J. N., \& Wolf, R. A., 1976, ApJ, 209, 214
\bibitem[1977]{BW77}Bahcall, J. N., \& Wolf, R. A., 1977, ApJ, 216, 883
\bibitem[1987]{BT87}Binney, J., \& Tremaine, S., 1987, Galactic Dynamics, (Princeton University
Press: Princeton), p. 541
\bibitem[1996]{Blum96}Blum, R. D., Sellgren, K., \& DePoy, D. L., 1996, ApJ, 470, 864
\bibitem[1995]{Carney95}Carney, B., Fulbright, J. P., Terndrup, D. M., Suntzeff, N., \& Walker, A.,
1995, AJ, 110, 1674
\bibitem[1998]{Davidge98}Davidge, T. J., 1998, AJ, 115, 2374
\bibitem[1997]{Davidge97}Davidge, T. J., Simons, D. A., Rigaut, F., Doyon, R., \& Crampton, D., 1997,
AJ, 114, 2586
\bibitem[1991]{DB91}Davies, M. B., \& Benz, W., 1991, ApJ, 381, 449
\bibitem[1998]{Davies98}Davies, M. B., Blackwell, R., Bailey, V. C., \& Sigurdsson, S., 1998, MNRAS,
301, 745
\bibitem[1997]{EG97}Eckart, A., \& Genzel, R., 1997, MNRAS, 284, 576
\bibitem[1993]{Eckart93}Eckart, A., Genzel, R., Hofmann, R., Sams, B. J. \& Tacconi-Garman, L. E., 1993,
ApJ, 407, L77
\bibitem[1995]{Eckart95}Eckart, A., Genzel, R., Hofmann, R., Sams, B. J. \& Tacconi-Garman, L. E., 1995,
ApJ, 445, L23
\bibitem[1976]{Frank76}Frank, J., \& Rees, M. J., 1976, MNRAS, 176, 633
\bibitem[1998]{Ghez98}Ghez, A. M., Klein, B. L., Morris, M., \& Becklin, E. E. 1998, ApJ, 509, 678
\bibitem[1994]{GHT94}Genzel, R., Hollenbach, D., \& Townes, C. H., 1994, Rep. Prog. Phys., 57, 417
\bibitem[1997]{Genzel97}Genzel, R., Eckart, A., Ott, T. \& Eisenhauer, F. 1997, MNRAS, 291, 21
\bibitem[1996]{Genzel96}Genzel, R., Thatte, N., Krabbe, A., Kroker H. \& Tacconi-Garman, L. E. 1996,
ApJ, 472, 153
\bibitem[1998]{Holtzman98}Holtzman, J. A., et al., 1998, AJ, 115, 1946
\bibitem[1982]{LTH82}Lacy, J. H., Townes, C H., \& Hollenbach, D. J., 1982, ApJ, 262, 120
\bibitem[1989]{LG89}Lee, M.-H., \& Goodman, J., 1989, ApJ, 343, 594
\bibitem[1988]{LT88}Livne, E., \& Tuchman, Y., 1988, ApJ, 332, 271
\bibitem[1992]{Kent92}Kent, S. M., 1992, ApJ, 387, 181
\bibitem[1995]{KR95}Kormendy, J., \& Richstone, D., 1995, AR\&A, 33, 581
\bibitem[1995]{Krabbe95}Krabbe, A., et al., 1995, ApJ, 447, L95
\bibitem[1998]{Magorrian98}Magorrian, J., et al., 1998, AJ, 115, 2285
\bibitem[1999]{MT99}Magorrian, J. \& Tremaine S., 1999, MNRAS, submitted (astro-ph/9902032)
\bibitem[1979]{MS79}Miller, G. E., \& Scalo, J. M., 1979, ApJS, 41, 513 
\bibitem[1996]{MS96}Morris, M., \& Serabyn, E., 1996, ARA\&A, 34, 645
\bibitem[1991]{MCD91}Murphy, B. W., Cohn, H. N., \& Durisen, R. H., 1991, ApJ, 370, 60
\bibitem[1998]{Ott98}Ott, T., Eckart, A., \& Genzel, R., 1998, ApJ submitted
\bibitem[1989]{Phinney89}Phinney, E. S., 1989, in The Center of the Galaxy, ed. M. Morris (Dordrecht:
Kluwer), 543
\bibitem[1995]{Quinlan95}Quinlan, G. D., Hernquist, L., \& Sigurdsson, S., 1995, ApJ, 440, 554
\bibitem[1990]{RS90}Rasio, F. A., \& Shapiro, S. L., 1990, ApJ, 354, 201
\bibitem[1991]{RS91}Rasio, F. A., \& Shapiro, S. L., 1991, ApJ, 377, 559
\bibitem[1988]{Rees88}Rees, M., 1988, Nature, 333, 523
\bibitem[1988]{Rieke88}Rieke, G. H., \& Rieke, M. J., 1988, ApJ, 330, L33
\bibitem[1986]{Scalo86}Scalo, J. M., 1986, Fundam. Cosmic. Phys., 11, 1
\bibitem[1993]{Schaerer93}Schaerer, D., Charbonnel, C., Meynet, G., Maeder, A., \& Schaller, G., 1993,
A\&AS, 102, 339
\bibitem[1990]{Sellgren90}Sellgren, K., McGinn, M. T., Becklin, E. E., \& Hall, D. N. B., 1990, ApJ, 359,
112
\bibitem[1996]{SM96}Serabyn, E., \& Morris, M. 1996, Nature, 382, 602
\bibitem[1998]{SSF98}Serabyn, E., Shupe, D., \& Figer, D. F. 1998, Nature, 394, 448
\bibitem[1971]{SH71}Spitzer, L. \& Hart, M. H., 1971, ApJ, 164, 399
\bibitem[1995]{Tiede95}Tiede, G. P., Frogel, J. A., \& Terndrup, D. M., 1995, AJ, 110, 2788
\bibitem[1980]{Young80}Young, P., 1980, ApJ, 242, 1232
\end{thebibliography}
\end{document}